\documentclass[fleqn,usenatbib]{mnras}

\usepackage{newtxtext,newtxmath}
\usepackage{multirow}
\usepackage{hyperref}
\usepackage{xcolor}
\usepackage{float}

\usepackage[T1]{fontenc}

\DeclareRobustCommand{\VAN}[3]{#2}
\let\VANthebibliography\thebibliography
\def\thebibliography{\DeclareRobustCommand{\VAN}[3]{##3}\VANthebibliography}

\usepackage{graphicx}	
\usepackage{amsmath}	
\usepackage[flushleft]{threeparttable}
\usepackage{float}

\usepackage{lineno}

\newcommand{\kms}{km~s$^{-1}$}
\newcommand{\sn}{SN~2021rhu}
\newcommand{\msun}{M$_\odot$}
\newcommand{\TiII} {\ion{Ti}{ii}}
\newcommand{\SiII} {\ion{Si}{ii}}
\newcommand{\CII} {\ion{C}{ii}}
\newcommand{\CI} {\ion{C}{i}}
\newcommand{\OI}{\ion{O}{i}}
\newcommand{\CaII}{\ion{Ca}{ii}}
\newcommand{\SII} {\ion{S}{ii}}
\newcommand{\MgII} {\ion{Mg}{ii}}

\title[Spectroscopic modelling of SN~2021rhu]{Early-time spectroscopic modelling of the transitional Type Ia Supernova 2021rhu with TARDIS}

\author[L. Harvey et al.]{
L. Harvey$^{1}$\thanks{E-mail: luharvey@tcd.ie},
K. Maguire$^{1}$,
M. R. Magee$^{2}$,
M. Bulla$^{3,4}$,
S. Dhawan$^{5}$,
S. Schulze$^{4}$,
J. Sollerman$^{4}$,
M. Deckers$^{1}$, \newauthor
G. Dimitriadis$^{1}$,
S. Reusch$^{6,7}$,
M. Smith$^{8}$,
J. Terwel$^{1,9}$,
M. W. Coughlin$^{10}$,
F. Masci$^{11}$,
J. Purdum$^{12}$,
A. Reedy$^{12}$, \newauthor
E. Robert$^{8}$,
and A. Wold$^{11}$
\\
$^{1}$School of Physics, Trinity College Dublin, The University of Dublin, Dublin 2, Ireland\\
$^{2}$Institute of Cosmology and Gravitation, University of Portsmouth, Burnaby Road, Portsmouth PO1 3FX, UK\\
$^{3}$Department of Physics and Earth Science, University of Ferrara, via Saragat 1, 44122 Ferrara, Italy\\
$^{4}$The Oskar Klein Centre, Department of Astronomy, Stockholm University, Albanova University Center, SE 106 91 Stockholm, Sweden\\
$^{5}$Institute of Astronomy and Kavli Institute for Cosmology, University of Cambridge, Madingley Road, Cambridge CB3 0HA, UK\\
$^{6}$Deutsches Elektronen Synchrotron DESY, Platanenallee 6, D-15738 Zeuthen, Germany\\
$^{7}$Institut für Physik, Humboldt-Universität zu Berlin, D-12489 Berlin, Germany\\
$^{8}$Université de Lyon, Université Claude Bernard Lyon 1, CNRS/IN2P3, IP2I Lyon, F-69622, Villeurbaan, France\\
$^{9}$Isaac Newton Group (ING), Apt. de correos 321, E-38700, Santa Cruz de La Palma, Canary Islands, Spain\\
$^{10}$School of Physics and Astronomy, University of Minnesota, Minneapolis, Minnesota 55455, USA\\
$^{11}$IPAC, California Institute of Technology, 1200 E. California Blvd, Pasadena, CA 91125, USA\\
$^{12}$Caltech Optical Observatories, California Institute of Technology, Pasadena, CA 91125, USA\\
}

\date{Accepted XXX. Received YYY; in original form ZZZ}

\pubyear{2021}

\begin{document}
\label{firstpage}
\pagerange{\pageref{firstpage}--\pageref{lastpage}}
\maketitle

\begin{abstract}
An open question in SN Ia research is where the boundary lies between `normal' Type Ia supernovae (SNe Ia) that are used in cosmological measurements and those that sit off the Phillips relation.  We present the spectroscopic modelling of one such `86G-like' transitional SN Ia, SN~2021rhu, that has recently been employed as a local Hubble Constant calibrator using a tip of the red-giant branch measurement. We detail its modelling from $-$12~d until maximum brightness using the radiative-transfer spectral-synthesis code \textsc{tardis}. We base our modelling on literature delayed-detonation and deflagration models of Chandrasekhar mass white dwarfs, as well as the double-detonation models of sub-Chandrasekhar mass white dwarfs. We present a new method for `projecting' abundance profiles to different density profiles for ease of computation. Due to the small velocity extent and low outer densities of the W7 profile, we find it inadequate to reproduce the evolution of SN~2021rhu as it fails to match the high-velocity calcium components. The host extinction of SN~2021rhu is uncertain but we use modelling with and without an extinction correction to set lower and upper limits on the abundances of individual species. Comparing these limits to literature models we conclude that the spectral evolution of SN~2021rhu is also incompatible with double-detonation scenarios, lying more in line with those resulting from the delayed-detonation mechanism (although there are some discrepancies, in particular a larger titanium abundance in SN~2021rhu compared to the literature). This suggests that SN~2021rhu is likely a lower luminosity, and hence lower temperature, version of a normal SN Ia.

\end{abstract}




Type Ia supernovae (SNe Ia) are the thermonuclear explosions of carbon-oxygen white dwarfs arising from interactions with a binary companion. Standardisable through the relationship between their peak luminosity and light curve width \citep[Phillips relation;][]{phillips_relation_pskovskii, phillips_relation_phillips} their use as distance indicators has been integral to the field of cosmology, leading to the discovery of the accelerating expansion of the Universe \citep{dark_energy_riess, dark_energy_perlmutter} and the theoretical prediction of dark energy. The power of SNe Ia as standardisable candles is predicated upon how strictly they follow the correlation between their peak luminosity and the speed of their light curve evolution. However, as the sample of observed SNe Ia has grown, it has become more diverse, with groups of transients clustering away from the Phillips relation \citep{Taubenberger_extremes}. Whether these outlying transients are produced by different progenitor channels  and/or explosion scenarios is still unclear \citep{ashley_review}. 

SN~1986G was the first of these outlier SNe Ia to cast doubt upon the usefulness of SNe Ia as standardisable candles \citep{1986G_observations, 1986G_modelling}. It was subluminous - with an absolute \textit{B}-band mag of $-$18.24$\pm$0.13 \citep{1986G_observations} - for its light curve decline in the $B$-band ($\Delta m_\text{15}$(B), the $B$-band magnitude decrease in the 15 days post peak). It also possessed a strong \TiII\ 4300 \AA\ absorption feature, which had not been seen in thermonuclear events up until this point. The years that followed brought the discovery of SN~1991bg, which was significantly fainter than SN 1986G, with a faster decline and a more pronounced titanium `trough’ \citep{91bg_spectrum, 1991bg_velocities}. In the decades since, many analogous objects have been discovered, with the subclass now labelled as the 91bg-like SNe. With this classification, SN~1986G is now considered to lie in the `transitional' region between the normal SNe Ia and the 91bg-like subclass.

It remains an open question as to whether transitional SNe Ia like SN\,1986G are suitable for use in cosmological measurements. Some transitional events have been used as distance ladder calibrator objects for H$_0$ measurements, with one transitional event, SN 2011iv \citep{2011iv}, being used as a calibrator SN for the recent SH0ES H$_0$ measurement \citep{SH0ES_H0}. SN~2011iv was also used as a calibrator object in the tip of the red giant branch (TRGB) calibration method \citep{2021Freedman,2022Anand}. In the case of \cite{SH0ES_H0}, the calibrator objects were required to pass cuts on SALT2 \citep{SALT2guy} colour ($|c|$ < 0.15) and stretch ($|x1|$ < 2) to be determined adequate for calibration, with another transitional object SN 2007on being excluded with $x1=-2.2$. However, SN 2007on was included in other H$_0$ measurements \citep[e.g.][]{2022Anand}.  More recently, the SN Ia, SN~2021rhu (ZTF21abiuvdk) was discovered by the Zwicky Transient Facility (ZTF) \citep{ztf1, ztf2, ztf3, ztf4} in a very nearby galaxy (NGC 7814) just three days after first light. Given its proximity to Earth and its location in a galaxy suitable for making TRGB measurements, a high cadence photometric and spectroscopic follow-up campaign of SN~2021rhu was triggered. The first results using it as a distance ladder calibrator object for a measurement of H$_0$ via the TRGB method were presented in \cite{Suhail_H0}. \cite{spectropolar} presented spectropolarimetric observations for SN~2021rhu in which they found a high degree of calcium polarisation 80~d post peak.

There are a large number of explosion models that have been proposed to explain both normal, transitional, and subluminous SNe Ia. One of the most popular model involving the explosion of a Chandrasekhar-mass white dwarf is the delayed-detonation model, where an initial sub-sonic deflagration phase transitions to a detonation \citep[e.g.][]{N100, n100_2}. The direct deflagration of a Chandrasekhar-mass white dwarf has also been studied but results in ejecta that are too mixed to be consistent with spectral observations of normal SNe Ia \citep{W7_density}. Sub-Chandrasekhar mass explosions have also been suggested to explain normal and sub-luminous SNe Ia. One such model is the double-detonation scenario, where a layer of He on the surface of the white dwarf explodes and subsequently triggers a secondary detonation of the core \citep{Nomotodoubledet,Livnedoubledet,Finkdoubledet,Shendoubledet,Sub_Ch_21rhu}. 

\cite{1986G_modelling} investigated the potential explosion mechanism and progenitor scenario for the original transitional event, SN~1986G, using modelling of its spectra. They concluded that the observed properties are most consistent with a low energy version of the deflagration of a Chandrasekhar mass white dwarf and that SN~1986G was therefore at the lower end of the `normal' SN Ia distribution (e.g.~originating from the same progenitor channel), instead of being part of the sub-luminous 91bg-like class. This supports the inclusion of transitional events in H$_0$ measurements.  However, the spectral series explored in the modelling of SN~1986G commenced 3 days before maximum light, and therefore this work could not probe the outer high-velocity region of the ejecta. It is noted that two earlier spectra were taken for SN~1986G; however, the limited wavelength range caused their exclusion in the modelling. Modelling of earlier spectra of a transitional event would allow us to place tighter constraints on the composition and densities of the higher velocity material, and in turn compare these transitional SNe Ia to literature explosion models and determine their link (or not) to normal cosmologically useful SNe Ia. 

In this paper, we study the very nearby and well observed transitional event, SN~2021rhu, that has been used in \cite{Suhail_H0} for a H$_0$ measurement. Our aim is to perform detailed radiative transfer modelling of its spectra from very soon after explosion,  to determine its ejecta structure and composition so that they can be linked to the most likely explosion scenario, and to determine its similarities (or deviations) from normal SNe Ia. In Section~\ref{observations}, we present the observational data and analysis for SN~2021rhu followed by the spectroscopic modelling method in Section~\ref{simulations} and the results in Section~\ref{results}. This is followed by a discussion of the models in the context of literature models in Section~\ref{discussion}, while the conclusions are presented in Section~\ref{conclusions}.

\section{Observations}
\label{observations}
\begin{figure}
 \includegraphics[width = \columnwidth]{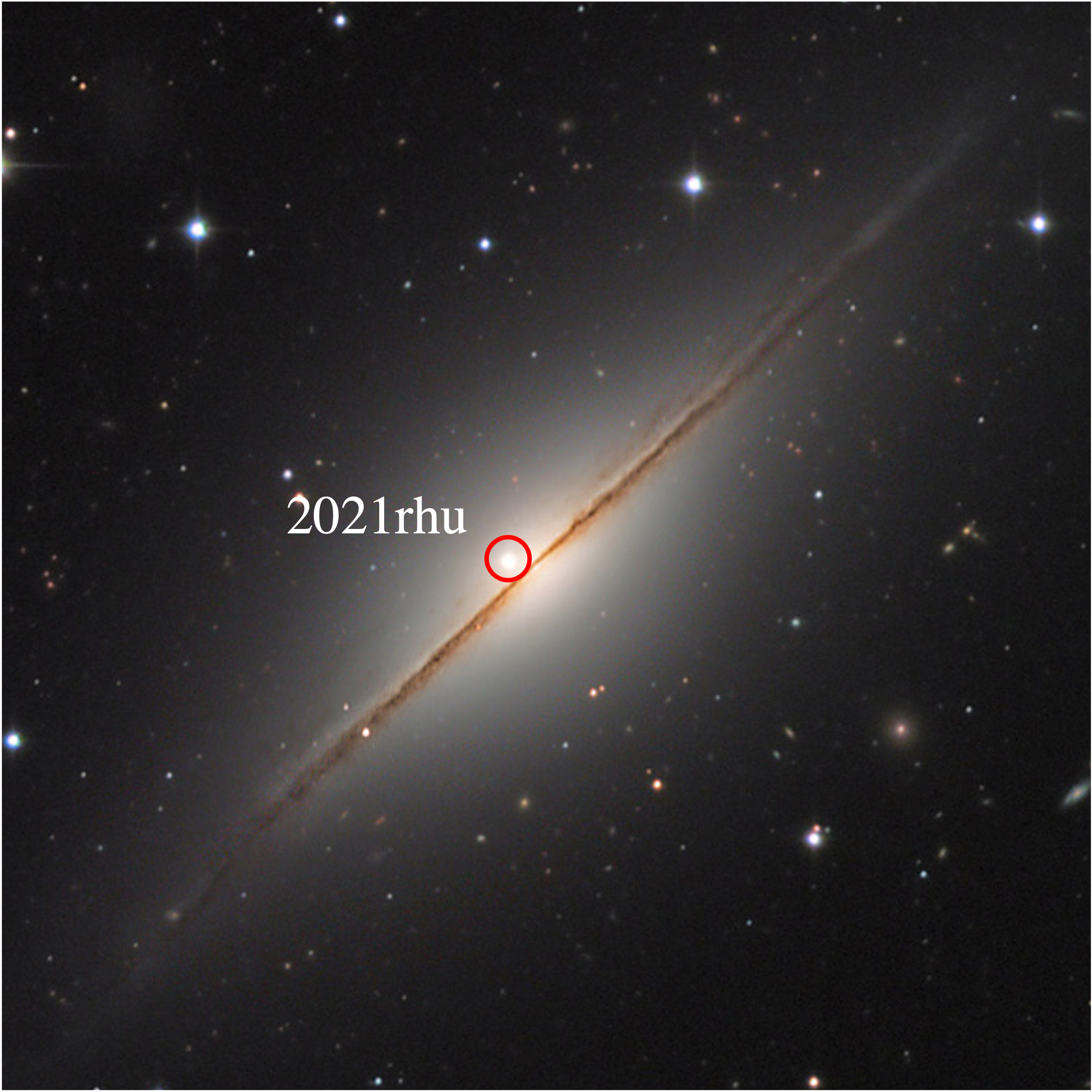}
 \caption{Host galaxy NGC 7814 showing SN~2021rhu in the red circle just outside the plane in the region of the bulge. \textit{Credit: Prompt 7 CTIO/UNC Chile, Chart32-Team.}}
 \label{fig:context_map}
\end{figure}

SN~2021rhu was discovered on 2021-07-01 (Modified Julian Date, MJD of 59396.56) at 15.66 mag in the ZTF~\textit{r}-band with the ZTF camera mounted on the 48-inch Samuel Oschin telescope (P48) at the Palomar observatory. It was announced to the TNS by the ALeRCE group \citep{ALeRCE}. At RA 00:03:15.42, Dec +16:08:44.51, this SN was found just outside the plane of the edge-on spiral galaxy NGC 7814 (Fig.~\ref{fig:context_map}) at a redshift of z = 0.003506 \citep{NGC7814}. The SN Ia classification came from the ALeRCE spectrum taken 2021-07-02, MJD 59396.94 \citep{2021rhu_classification} with the Transient Double-beam Spectrograph on the 2.5m telescope of the Caucasus Mountain Observatory \citep{DBS-CMO}. Identified as an interesting object due to its low redshift and early discovery, an extensive monitoring campaign was initiated. 

\subsection{Distance and extinction towards SN~2021rhu}
Due to its proximity the Earth, SN~2021rhu does not lie in the Hubble flow, and as such a redshift-independent distance is required for correcting photometry to absolute magnitudes and scaling the spectra from flux to luminosity. A range of distance measurements exists from previous works for the host galaxy of SN~2021rhu. Here we adopt the updated TRGB measurement of $\mu=30.86\pm0.07$~mag from \cite{Suhail_H0}, where SN~2021rhu was used as a calibrator object for a H$_0$ measurement.

The SN~2021rhu photometry was corrected for Milky Way extinction in accordance with the dust extinction model from \cite{fm07} with $R_\text{v}=3.1$ and $E(B-V)=0.0383\pm0.0003$~mag, as taken from \cite{IRSA_dustmap} accessed through the \textsc{python} package \textsc{astroquery} \citep{astroquery}.

\begin{figure}
 \includegraphics[width = \columnwidth]{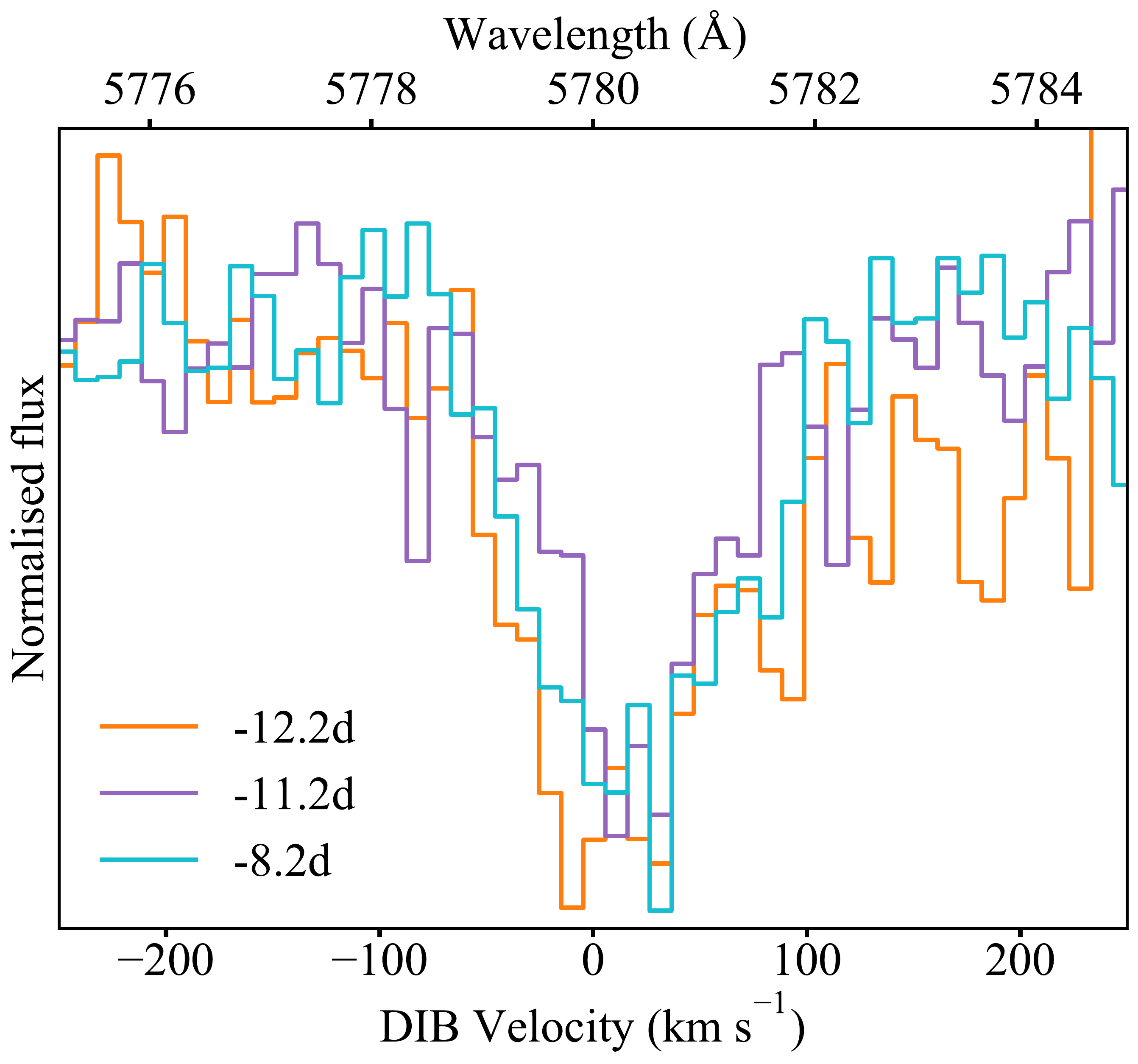}
 \caption{The diffuse interstellar band in the intermediate-resolution XShooter spectra at epochs of $-$12.2~d, $-$11.2~d, and $-$8.2~d with respect to peak. The equivalent width measurement of this feature gives an estimate of the extent to which the data is affected by host extinction.}
 \label{fig:DIB}
\end{figure}

With three intermediate resolution XShooter spectra - further discussed in Section~\ref{sec:spectroscopy} - we are able to measure the equivalent width of the diffuse interstellar band at 5780 Å (see Fig.~\ref{fig:DIB}) which has been shown to display a correlation with host $A_\text{v}$ by \cite{DIB_extinction}. While the relation has large uncertainties, with a 1$\sigma$ dispersion of 50 per cent, it has been shown to be a significantly better indicator of host extinction than the more commonly used Na I D lines \citep{DIB_extinction}. We measure a value of 80$\pm$1 mÅ, which corresponds to a host extinction of $A_\text{v}=0.42\pm0.21$ mag. Lying just out of the plane of the host galaxy NGC 7814, as seen in Fig.~\ref{fig:context_map}, this potentially large host extinction is unsurprising. Since there is a large uncertainty on the host extinction value, we have explored radiative transfer models with and without this host extinction correction applied (see Section \ref{simulations}). These two models can then be used to impose upper and lower limits on elemental abundances.

\subsection{Photometry}
\label{sec:observations}
After discovery, daily ZTF~\textit{gri} photometry was obtained on the P48 telescope, leading to very high cadence light curves in the three ZTF bands over the evolution of SN~2021rhu. The photometric data presented here is from the \textsc{ztffps} forced photometry pipeline \citep{ztffps}.

We observed the field with the 30 cm Ultraviolet/Optical Telescope \citep[UVOT;][]{Roming2005a} aboard the \textit{Swift} satellite \citep{Gehrels2004a} between 2021-07-15 and 2021-07-25 in $w2$, $m2$, $w1$, $u$, $b$, $v$ bands with a $\sim$3d cadence. After the SN faded, we obtained a final set of images in September 2022 to remove the host contribution. Data were retrieved from the NASA Swift Data Archive \footnote{ \url{https://heasarc.gsfc.nasa.gov/cgi-bin/W3Browse/swift.pl}} and processed using UVOT data analysis software HEASoft version 6.30.1\footnote{ \url{https://heasarc.gsfc.nasa.gov/}}. Source counts were extracted from the images using a region of $5''$. The background was estimated using a significantly larger region outside of the host galaxy. The count rates were obtained from the images using the \textit{Swift} tool \textsc{uvotsource}. They were converted to magnitudes using the UVOT photometric zero points \citep{Breeveld2011a} and the latest calibration files from February 2022. To remove the host emission from the transient light curves, we used templates formed from our final observations in September 2022. We measured the host contribution using the same source and background apertures and subtracted this contribution from the transient flux measurements. Unfortunately the target was saturated in the $b$-band images and as such this data is excluded. The observed photometric data for SN~2021rhu can be found in Table~\ref{tab:photometry}. K corrections have not been applied due to the low redshift of the host galaxy.

\begin{figure}
 \includegraphics[width = \columnwidth]{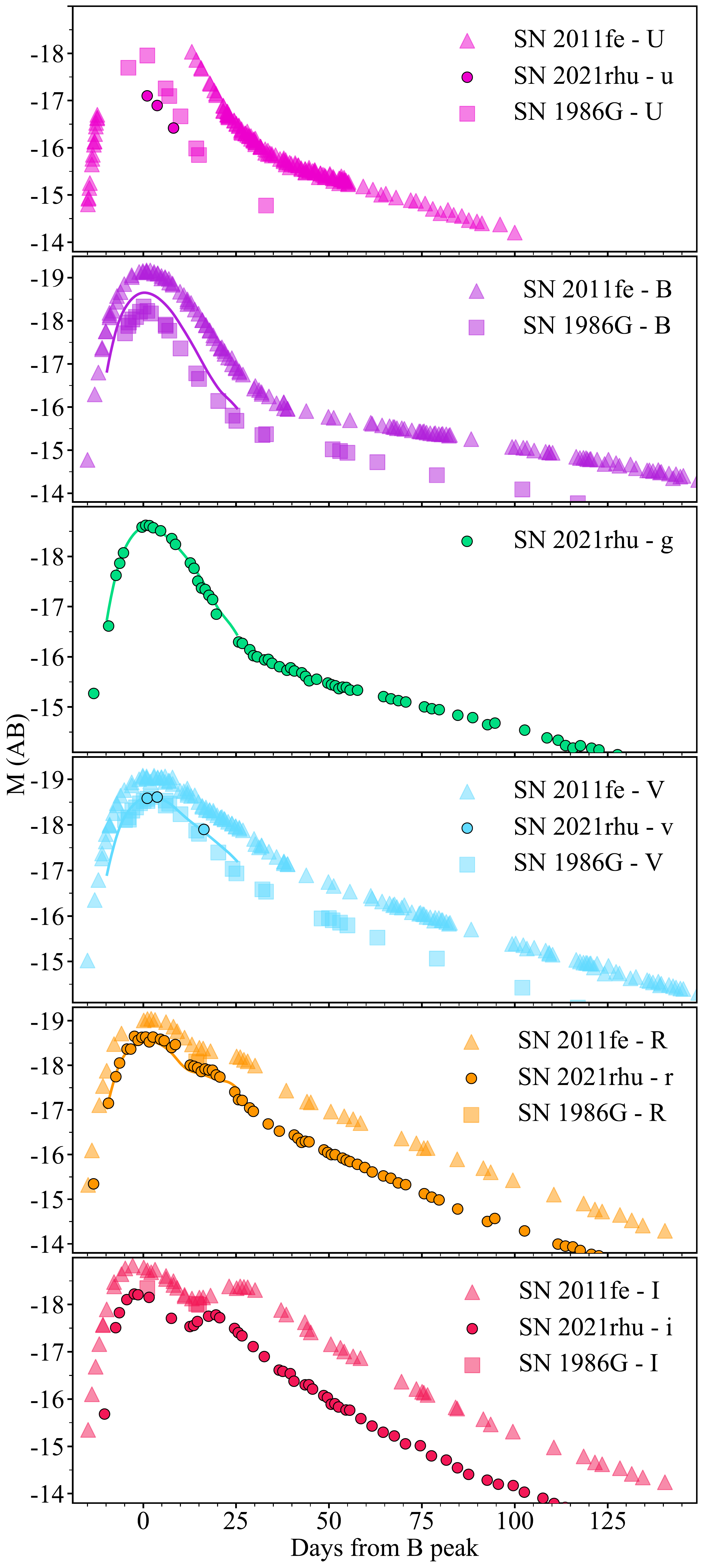}
 \caption{Absolute magnitude light curves with no host extinction applied for SN~2021rhu in the \textit{Swift} \textit{u} and \textit{v} bands and the ZTF~\textit{g}, ZTF~\textit{r}, and ZTF~\textit{i} bands compared with the \textit{UBVRI} photometry of transitional event, SN~1986G and the normal SN~2011fe. The SALT2 light curve fits for SN~2021rhu are displayed as solid lines. We note that the fitting was performed on only the ZTF~\textit{gr} data and as such the solid lines in the BV bands are extrapolations. The extrapolation from this SALT fit was left out in the case of the $U$ and $I$ bands as these regions are known to be weaknesses of SALT2. The extrapolated $B$-band light-curve is the source of the peak $M_\text{B}$ and $\Delta m_\text{15}$(B) measurements for SN~2021rhu seen in Fig.~\ref{fig:Taubenberger}.}
 \label{fig:LCs}
\end{figure}

The absolute magnitude light curves of SN~2021rhu are shown in Fig.~\ref{fig:LCs}  along with the \textit{UBVRI} photometry for the transitional event, SN~1986G \citep{1986G_observations} and for the well studied normal Type Ia SN~2011fe \citep{2011fe_photometry, 11fe_BV, 11fe_U}.  The photometric data for SN~1986G and SN~2011fe were retrieved through the Open Supernova Catalog \citep{open_supernova_catalog}. We have corrected both their light curves for MW extinction and converted them to absolute magnitudes (see Table \ref{tab:correction_values} for values and their associated references). Host extinction for SN~1986G must also be accounted for as it is known to have occurred in a dust lane of its host galaxy Centaurus A \citep{1986G_observations}. For SN~1986G we adopt the host $A_\text{v}$ (see values in \ref{tab:correction_values}) from \cite{DIB_extinction} in which they measured column densities of neutral sodium and potassium.

\begin{figure}
 \includegraphics[width = \columnwidth]{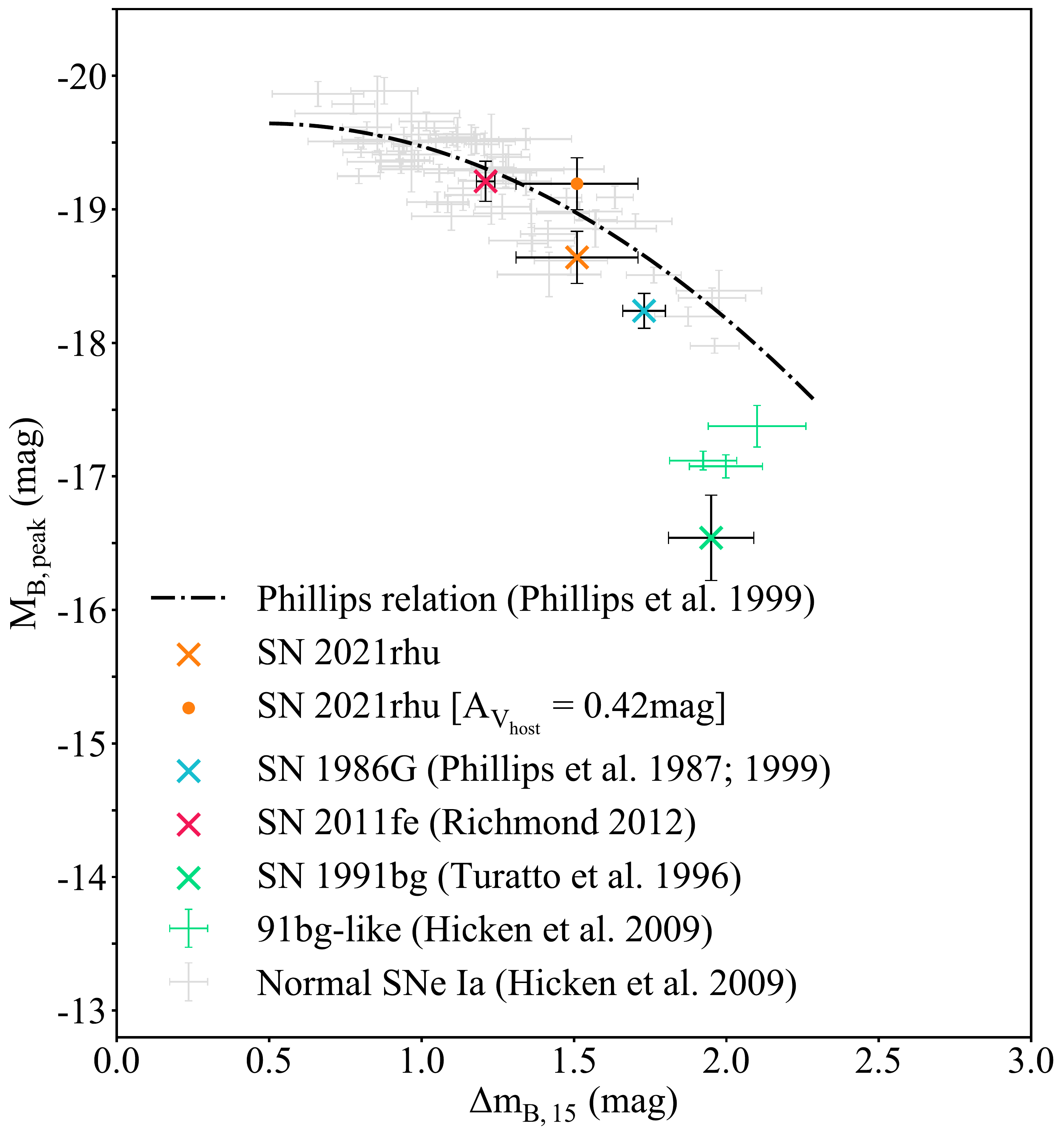}
 \caption{Locations of SN~1991bg, SN~1986G, SN~2021rhu, and SN~2011fe in the parameter space of the Phillips relation. As evident from its intermediate location between SN~2011fe and the transitional SN~1986G, SN~2021rhu lies below the Phillips relation in a similar region to SN~1986G. The Phillips relation is represented by the black line \citep{Phillips99}.}
 \label{fig:Taubenberger}
\end{figure}

\subsection{Light-curve properties}
\label{sec:LC_properties}
As evident in the \textit{BV} photometry in Fig.~\ref{fig:LCs}, SN~1986G is subluminous when compared to normal SNe Ia. This also appears to be the case for SN~2021rhu as seen in the comparisons between its ZTF~\textit{ri} photometry with the \textit{RI} photometry available for SN~2011fe. We note that although the response functions of the ZTF~\textit{r} and ZTF~\textit{i} bands differ slightly from the \textit{RI} filters, this only leads to a magnitude difference on the order of 0.05 mag and as such  this luminosity comparison is still significant. 

The SN~2021rhu photometry was fit by \cite{Suhail_H0} using the light-curve fitter SALT2 \citep{salt} as accessed through \texttt{sncosmo} \citep{sncosmo}. This fit was performed solely on the ZTF~\textit{g} and ZTF~\textit{r} data as SALT2 is not well defined at wavelengths redward of 7000~Å. The date of maximum light was found to be $t_0=59410.60\pm0.04$. With SALT2 parameters of $x_1 = -2.074\pm0.025$ and $c = 0.054\pm0.028$, SN~2021rhu satisfied the criteria of $|x_1|<3$ and $|c|<0.3$ as laid out in \cite{Suhail_H0} as typical cosmological cuts. These criteria are slightly looser than the previously mentioned cuts employed in \cite{SH0ES_H0}. SN~2021rhu would have in fact been excluded as a calibrator for the SH0ES measurement. It was argued that despite the low $x_1$ value - also seen in other peculiar, fast-decliners - the clear ZTF~\textit{r} shoulder and ZTF~\textit{i} secondary maximum are characteristic of normal and transitional SNe Ia that have been used for cosmology. While there is only limited data available for SN~1986G in the \textit{I} and \textit{R} bands, it has a clear secondary maximum in the infrared photometry presented in \cite{frogel_86G_ir}. \cite{Suhail_H0} went a step further by calculating the colour-stretch parameter through another light-curve fitter, SNooPY as $s_{BV}=0.72$, which is consistent with normal and transitional thermonuclear events.

The resulting time-varying SED from the SALT2 fit was then integrated through the Bessel filter responses to produce the extrapolated light-curves for the $V$ and $B$ bands, seen as solid lines in Fig.~\ref{fig:LCs}. While SN~2021rhu appears subluminous in comparison to SN~2011fe, it also exhibits a faster decline and as such it is necessary to examine where it lies in the parameter space of the Phillips relation in order to determine the extent to which it strays from other thermonuclear events.

From the extrapolated $B$ light-curve taken from the SALT2 fit we measure the peak magnitude as $M_B = -18.64\pm0.2$ with $\Delta m_\text{15}$(B) = 1.51$\pm0.2$ mag.  As seen in Fig.~\ref{fig:Taubenberger}, these values put SN~2021rhu below the Phillips relation, in the subluminous region of the parameter space. At a similar distance from the luminosity-width relation as its faster evolving analogue SN~1986G, SN~2021rhu appears to be a transitional transient.  If we include the $A_\text{v}=0.42\pm0.21$ mag host extinction measurement, SN~2021rhu would be positioned just above the Phillips relation (Fig.~\ref{fig:Taubenberger}). As this strong \TiII\ trough has only been previously seen in sub-luminous objects, we take this value as the host extinction upper limit. 

SALT2 is trained upon normal SNe Ia which do not possess this strong \TiII\ feature present in SN~2021rhu. This feature falls in the wavelength range of the $B$-band and as a result, measurements from this SALT extrapolation might overestimate the brightness of SN~2021rhu in the $B$-band. This overestimate shifts the tranisent upwards in Fig.~\ref{fig:Taubenberger} and as such it likely sits slightly lower in the parameter space, closer to SN~1986G.

\begin{figure}
 \includegraphics[width = \linewidth]{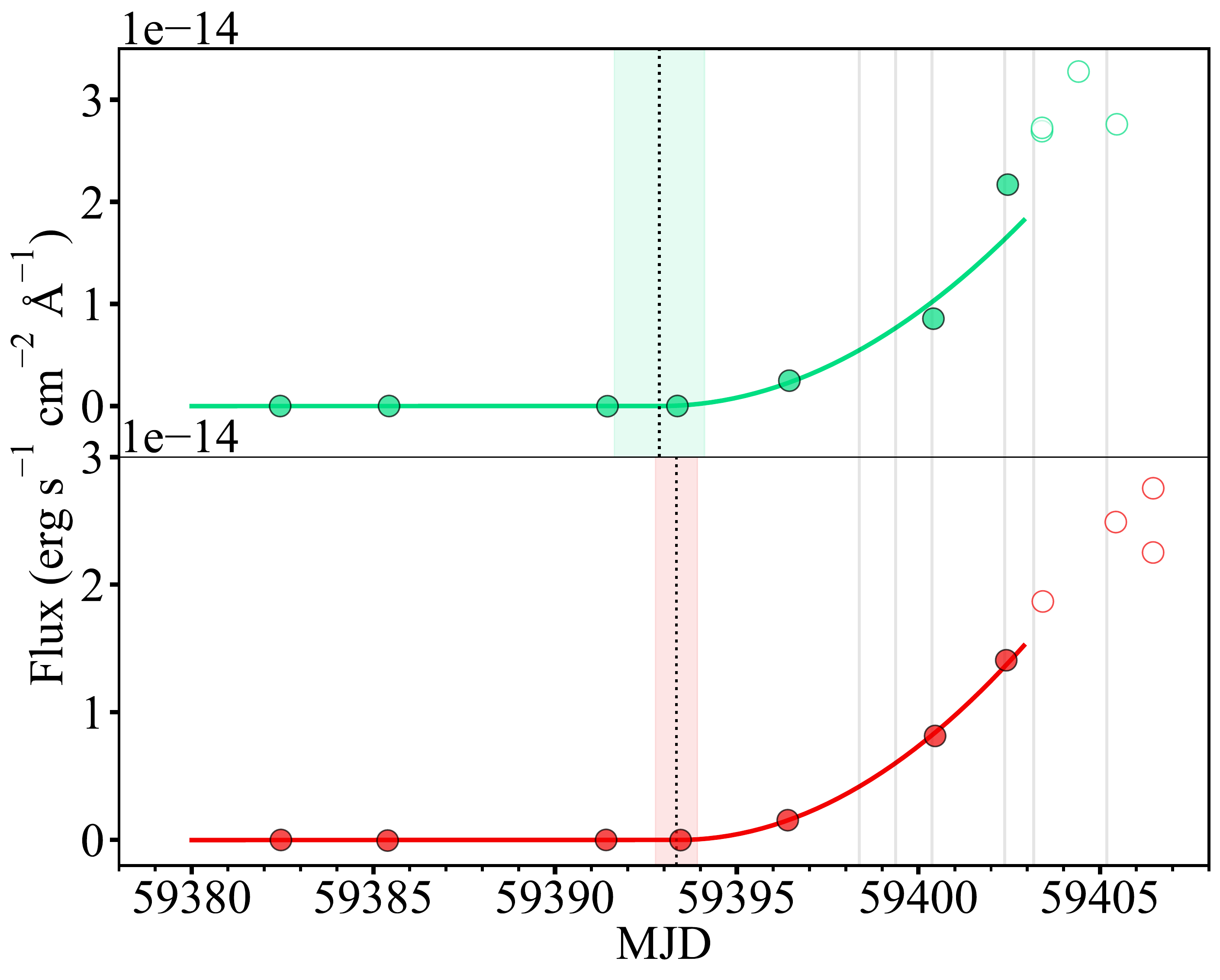}
 \caption{Results of the fitting of the ZTF~\textit{g} (top) and ZTF~\textit{r} (bottom) light-curve rises with Equation~\ref{eqn:early_LC_fit}. The filled data points are those included in the fit, while the open circles are not. The fitted date of first light is indicated by the black dotted lines with the shaded regions representing the 1$\sigma$ uncertainties. The grey vertical lines mark the dates of spectral epochs.}
 \label{fig:early_LC_fit}
\end{figure}

To obtain a date of first light, we employed the prescription from \cite{early_lc_fit} in which the early light-curve rise up to $\sim40$ per cent of the peak magnitude can be fit in flux space with the following power law:

\begin{equation}
\centering
    f(t) = C+H[t_{\text{fl}}]A(t-t_{\text{fl}})^{\alpha}.
\label{eqn:early_LC_fit}
\end{equation}

where $t_{\text{fl}}$ is the time of first light, $\alpha$ is the power law index, $A$ is the proportionality constant, $C$ is a constant offset, and $H[t_{\text{fl}}]$ is the Heaviside function taking the value $0$ before $t=t_{\text{fl}}$ and $1$ afterwards. The $40$ per cent cutoff as used by \cite{early_lc_fit} was noted to be an arbitrary choice and to ensure sufficient data for the fit, we included datapoints up to $50$ per cent of the peak flux.The date for first light was fit to be 59392.87$\pm$1.24~d in ZTF~\textit{g} and 59393.34$\pm$0.58~d in ZTF~\textit{r}. Combining these results with maximum light dates from the SALT fit for each band gives rise times of 17.99$\pm$1.24~d and 18.08$\pm$0.58~d in ZTF~\textit{g} and ZTF~\text{r} respectively. The fits to these two bands can be seen in Fig.~\ref{fig:early_LC_fit}. In previous studies the power law index has been fixed as $\alpha=2$ with the luminosity scaling with the expanding fireball surface area. When left as a free parameter, the resulting value can provide insight into the distribution of $^{56}$Ni. Smaller values of $\alpha$ describe a more gradual rise and point towards a more extended $^{56}$Ni distribution with a shorter time between explosion and first light (dark phase), whereas a large index implies a lesser degree of mixing with the $^{56}$Ni restricted to the core and all the photons escaping over the course of a smaller time frame following a longer dark phase \citep{Firth_lc_rise}. The fitted values of $\alpha$ were $2.00\pm0.58$ for ZTF~\textit{g} and $2.01\pm0.25$ for ZTF~\textit{r}.

\subsection{Spectroscopy}
\label{sec:spectroscopy}
Spectra were obtained for SN~2021rhu from $-$12 to $+$200 d with respect to peak brightness at a number of facilities: the XShooter spectrograph \citep{XSHOOTER} on the ESO Very Large Telescope (VLT) at the Paranal Observatory, the Spectral Energy Distribution Machine \citep[SEDM;][]{SEDM} on the automated P60 \citep{P60} at the Palomar Observatory, the Spectrograph for the Rapid Acquisition of Transients \citep[SPRAT;][]{SPRAT} at the Liverpool Telescope \citep[LT;][]{LT}, and the Alhambra Faint Object Spectrograph and Camera (ALFOSC)\footnote{\href{http://www.not.iac.es/instruments/alfosc}{{http://www.not.iac.es/instruments/alfosc}}} on the 2.56\,m Nordic Optical Telescope (NOT) at the Observatorio del Roque de los Muchachos on La Palma (Spain).  The observational details for the spectra are given in Fig.~\ref{tab:spectral_info_2021rhu}.

Data reduction for the XShooter spectra were performed following the method in \cite{Kate_XShooter} that uses the \textsc{REFLEX} pipeline \citep{XShooter_1, XShooter_2}. The SEDM spectra were reduced through the IFU pipeline developed by \cite{Rigault_SEDM} and \cite{Young-Lo_SEDM}. The SPRAT spectra were reduced through \textsc{pyraf} using a custom \textsc{python} script following \cite{Simon_LT}. The ALFOSC data were reduced with the python data reduction pipeline PyNOT\footnote{\href{https://github.com/jkrogager/PyNOT}{https://github.com/jkrogager/PyNOT}} in a standard fashion.

Absolute flux calibration of the spectra was carried out using the ZTF~\textit{g} and ZTF~\textit{r} bands as not only do they have a higher cadence than the ZTF~\textit{i} photometry, but many of the spectra do not possess a wavelength range that covers the ZTF~\textit{i} filter response function in its entirety. Synthetic photometry measurements were calculated from the spectra using the \textsc{pyphot} package \citep{pyphot} and then compared to the photometry values - if coinciding with the epoch of the spectrum - or an interpolated light curve. This interpolated light curve was comprised of a power law fit for the early rise and the SALT2 fit for the main body of the light curve. These comparisons resulted in calibration factors for each of the two bands at their corresponding effective wavelengths. Fitting a linear function to these calibration factors resulted in a wavelength dependant calibration function, which could then be applied to calibrate the spectra to the photometry. We note that this style of calibration can cause scaling issues with the spectrum in regions far away from the effective wavelengths of the filters. However, as these filters cover almost the entire range in which we are interested, this has no significant impact on the spectra presented here.

The spectra are subsequently corrected to the restframe using the redshift of the host galaxy, for extinction using the MW values, and to luminosity using the TRGB distance modulus, all given in Table~\ref{tab:correction_values}. The modelling of the spectra corrected for potential host galaxy extinction are discussed in Sections \ref{sec:ion_projection} and \ref{sec:N100*h}.

\begin{figure*}
 \includegraphics[width = \linewidth]{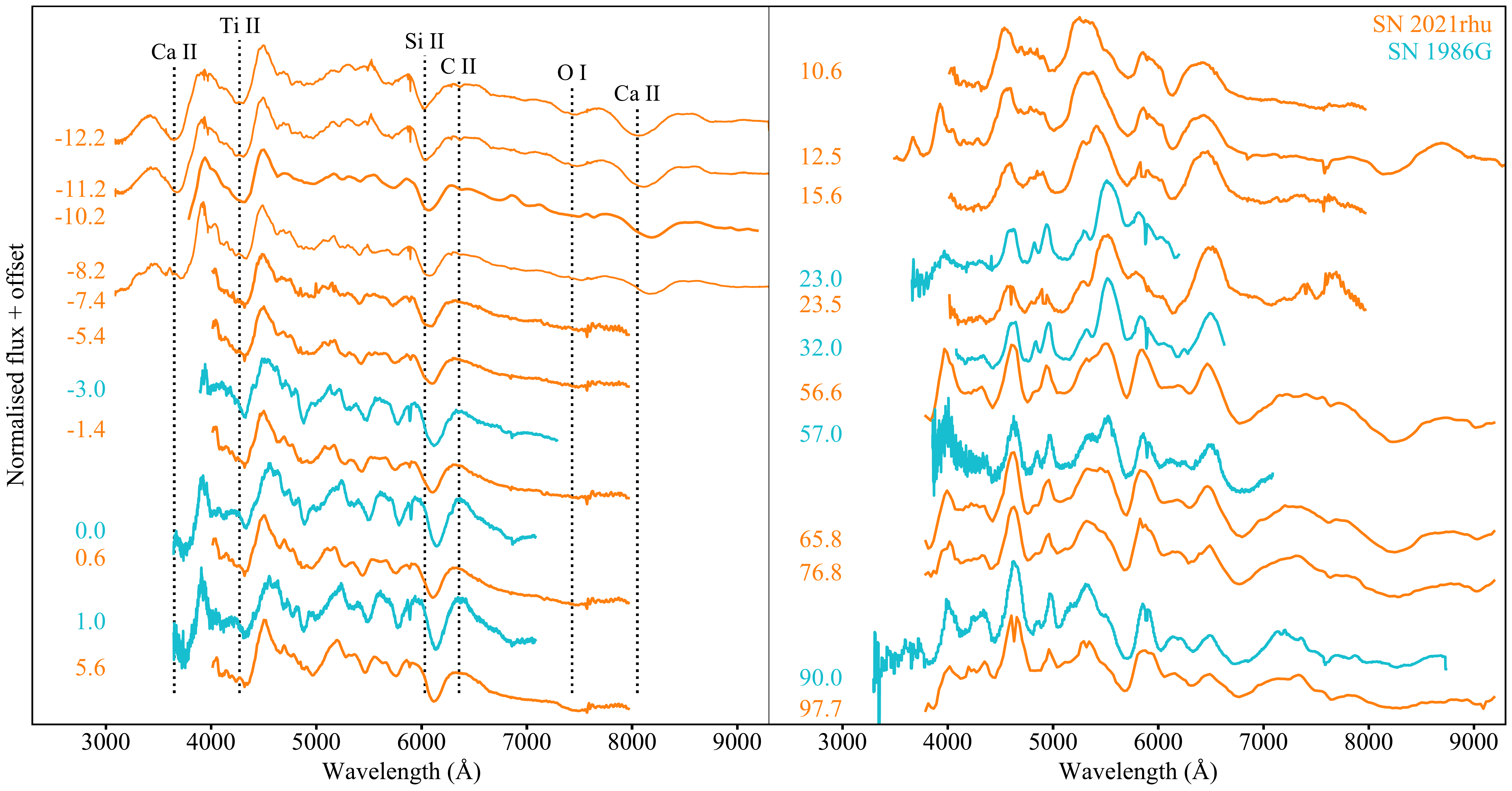}
 \caption{Spectral data for SN~2021rhu (orange) compared against the spectral evolution of SN~1986G (blue). The phase of each spectrum in days relative to peak is displayed alongside each spectrum, with some of the key features indicated by the vertical dotted lines. All spectra have been corrected for redshift and MW extinction. In the case of SN~1986G this extinction correction also included a component to compensate for the host dust lane. }
 \label{fig:spectral_comp}
\end{figure*}

\begin{figure}
 \includegraphics[width = 8cm]{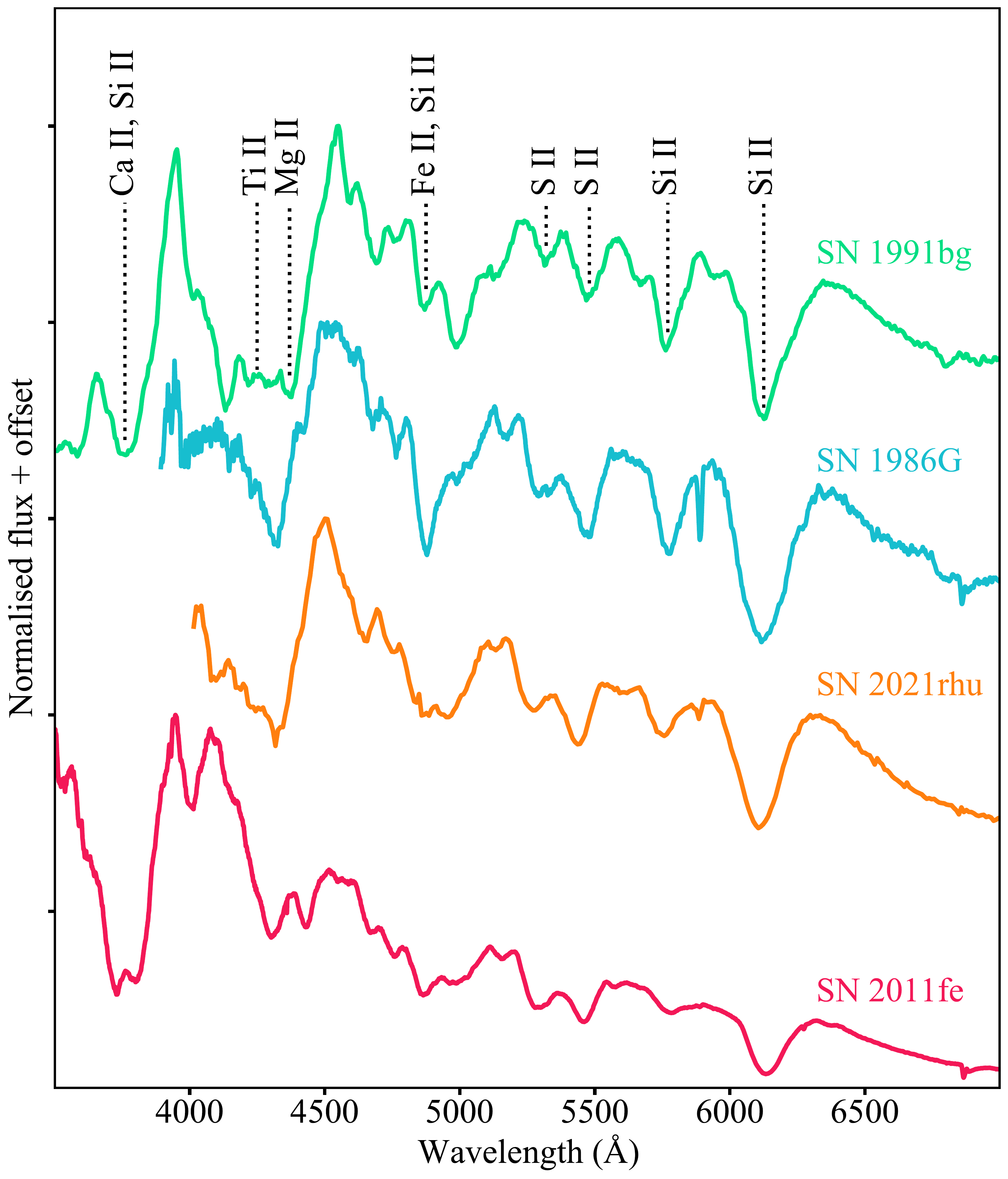}
 \caption{Comparison between the peak time spectra of SN~1991bg, SN~1986G, SN~2021rhu and SN~2011fe. Key lines are labelled along with the strong titanium feature at $\sim$4300~Å seen in SN~1991bg, SN~1986G, and SN~2021rhu that does not appear in the normal SNe Ia such as SN~2011fe.}
 \label{fig:peak_spectral_comp}
\end{figure}

\subsection{Observed spectral properties}
In Fig.~\ref{fig:spectral_comp} we present the spectral series for SN~2021rhu, shown alongside some of the available spectra for SN~1986G \citep{1986G_spectra} with similar epochs for comparison. The spectra of SN~2021rhu have only been corrected for MW extinction. In such a side-by-side comparison the resemblance between the two transients can be seen, with the only notable difference between the two being a slightly lower \SiII\ $\lambda6355$ velocity in SN~1986G. 

Figure~\ref{fig:peak_spectral_comp} compares the maximum-light spectrum of SN~2021rhu to those of SN~2011fe, SN~1986G, and SN~1991bg - the namesake of the subluminous 91bg-like subclass of thermonuclear transients \citep{91bg_spectrum, 1991bg_velocities, 91bg_phillips}. The \TiII\ feature seen in the SN~1991bg spectrum at $\sim$4300~Å is common to the subclass and can also be seen clearly in SN~1986G and SN~2021rhu although to a less pronounced extent. This feature is absent in all normal SNe Ia, such as SN~2011fe \citep{11fe_spectrum}. The presence of this \TiII\ absorption feature and stronger \SiII\ $\lambda5972$ absorption, along with a peak brightness between normal and 91bg-like SNe Ia, gives SN~1986G and SN~2021rhu their `transitional' classification.

\begin{figure}
 \includegraphics[width = \columnwidth]{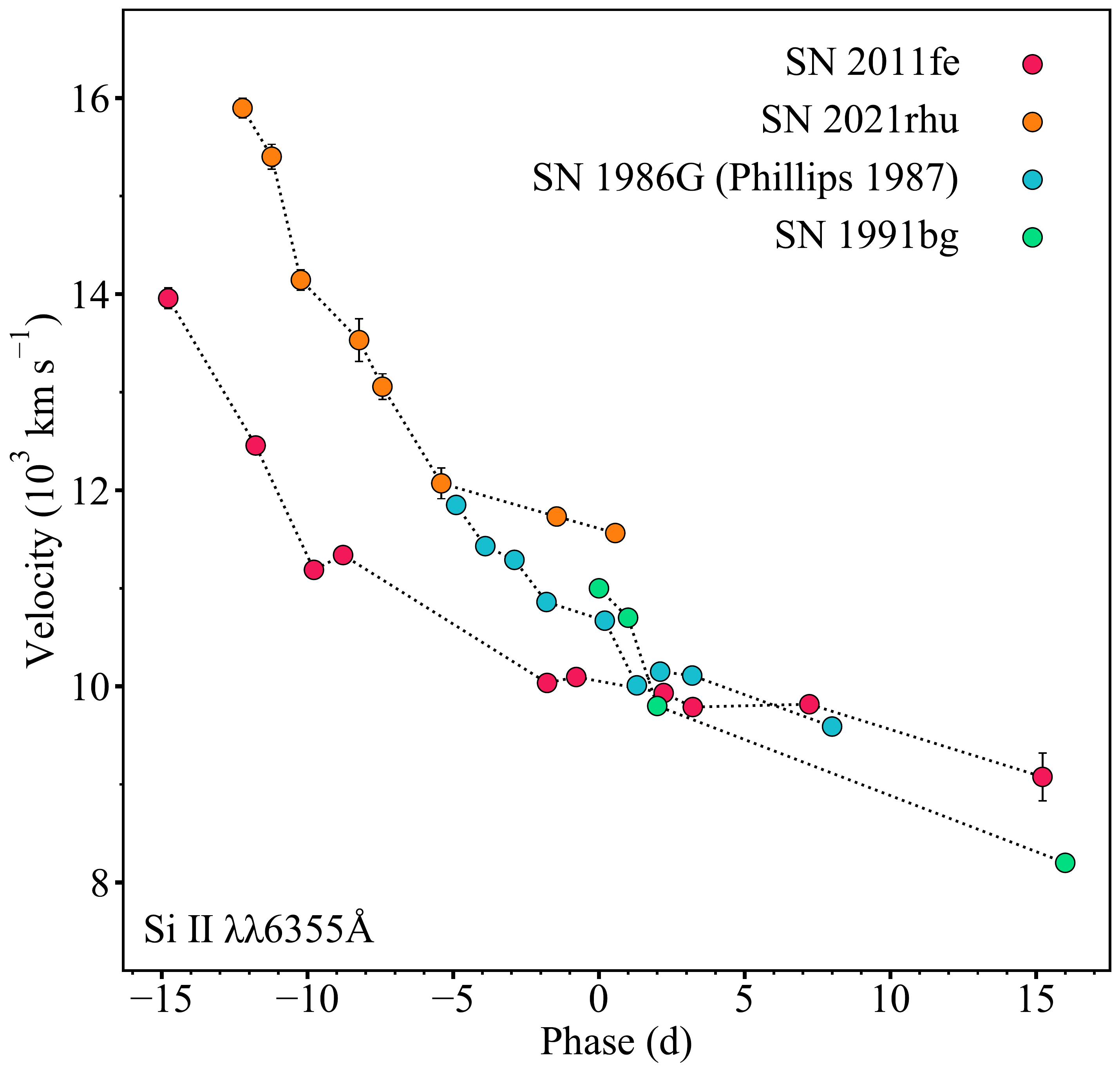}
 \caption{Velocity evolution of the \SiII\ $\lambda6355$ feature measured here for SN~2021rhu, SN~2011fe \citep{11fe_spectrum}, and SN~1991bg \citep{91bg_spectrum, 91bg_phillips} over the rise time, compared against the measurements of this line velocity for SN~1986G \citep{1986G_observations}.}
 \label{fig:velocity_evo}
\end{figure}

The velocity evolution of the \SiII\ $\lambda6355$ feature is measured for SN~1991bg, SN~2011fe, and \sn\ using Gaussian fits to the line profiles and varying the continuum positioning to estimate the associated uncertainties. The results of this line fitting is seen in Fig.~\ref{fig:velocity_evo}, along with the values measured for SN~1986G from \cite{1986G_observations}. The \SiII\ $\lambda6355$ for SN 2021rhu is $\sim$11500~\kms\ at peak brightness with a decline from $\sim$16000 \kms\ at $-$12~d. In the overlapping region with SN~1986G around peak, the \sn\ \SiII\ velocities are slightly higher than for SN~1986G velocities but are roughly consistent.
However, the SN~2021rhu \SiII\ velocities are consistently higher by $\sim2000$~\kms\ than the velocities seen in SN~2011fe at similar epochs. The mean \SiII\ velocity for SNe Ia around peak ($-$5 to +5 d) is $\sim$11500 \kms, with values typically in the range of $\sim$9500 -- 14000 \kms\ \citep{Kate_PTFspec}. Therefore, the peak \SiII\ velocity of \sn\ sits in the normal SN Ia range. 

A small feature to the red of the \SiII\ $\lambda6355$ is visible in all three intermediate-resolution XShooter spectra at $-$12.2, $-$11.2 and $-8.2$ d - as well as in the $-$10.2~d SEDM spectrum. The feature in this region is typically associated with carbon, specifically the \CII\ $\lambda6580$ line. If this feature is formed by \CII\, it resides at $\sim10000$~\kms\ at $-$12.2~d which is far below the $15898\pm100$~\kms\ velocity of the \SiII\ $\lambda6355$ line at this epoch. As the \SiII\ velocity typically traces the rough velocity evolution of the photosphere in the pre-peak regime, this feature - if formed by \CII\ $\lambda6580$ - exists at velocities some $6000$~\kms\ slower than the photosphere. The \CII\ $\lambda7236$ feature also appears to be present with a slightly higher velocity of $\sim11000$~\kms\ in the $-$12.2~d spectrum.

\begin{figure}
 \includegraphics[width = \linewidth]{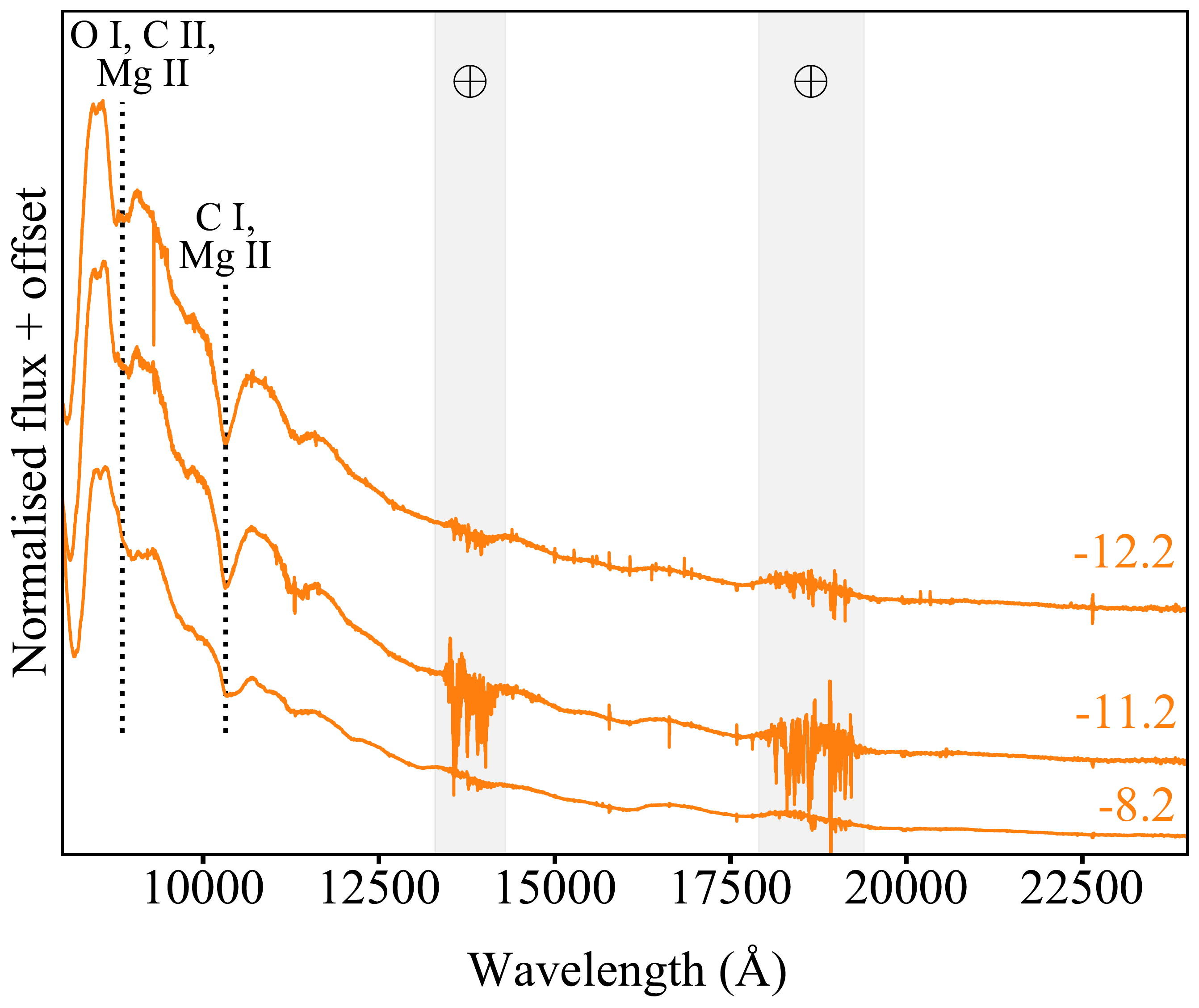}
 \caption{The infrared region of the three early XShooter epochs for SN~2021rhu. Once again key features are indicated by vertical dotted lines, with the grey shaded bands indicating the telluric regions.}
 \label{fig:spectral_comp_ir}
\end{figure}

Figure~\ref{fig:spectral_comp_ir} shows the near-infrared regions of the three XShooter spectra with the telluric regions indicated by the grey shaded bands. The strong feature here in the blue is the \CaII\ NIR as labelled in Fig.~\ref{fig:spectral_comp}. The feature directly to the red of this aligns with wavelengths of the \MgII\ $\lambda$9231 doublet, and the \OI\ $\lambda$9263 and \CII\ $\lambda$9234 lines shifted to velocities similar to the \SiII\ $\lambda$6355 feature at this epoch. The resulting feature is likely a blend of the three. Finally the strong feature at $\sim10200$Å falls in line with the \MgII\ 10927Å with possible contributions from \CI\ 10693 Å.

\section{Spectral modelling}
\label{simulations}
The aim of our spectral modelling for SN~2021rhu is to constrain the abundances of key elements through abundance tomography, which provide clues as to the explosion mechanism at play. We are particularly interested in the abundance of titanium required to produce the observed evolution of the strong feature seen in the spectra at $\sim$4300~Å. Due to the enhanced production of titanium, the double-detonation mechanism \citep{Shendoubledet,Sub_Ch_21rhu} is an interesting avenue of investigation in the context of lower luminosity thermonuclear events with strong \TiII\ absorption. It has also been discussed, however, that instead of higher titanium production, lower ejecta temperatures may be responsible for this increased absorption \citep{Mazzali_low_temp_Ti, Nugent_low_temp_Ti}, with this feature weakening at higher ejecta temperatures as most of the titanium is hidden away in higher ionisation states. 

The key difference between the previous spectral modelling work of \cite{1986G_modelling} for another transitional SN~1986G and the modelling discussed here is the ability of the \sn\ data to constrain the higher velocity material. The earliest spectrum explored for SN~1986G was recorded just 3~d before maximum light in comparison to 12.2~d before for SN~2021rhu. With five additional SN~2021rhu spectra taken in this $\sim$9d window we are able to better constrain the composition of the material in the regions of the ejecta above $\sim15000$~km~s$^{-1}$. In this Section, we describe the \textsc{tardis} code used to model the spectra of \sn, present the density profiles which we shall be exploring, introduce a number of methods to manipulate models to different density profiles and input parameters, as well as discuss the impact of including host extinction on the simulated spectra.

\subsection{\textsc{tardis}}
\label{TARDIS}
\textsc{tardis} is an open-source Monte-Carlo radiative-transfer spectral synthesis code for one-dimensional models of transients \citep{2014MNRAS.440..387K, kerzendorf_wolfgang_2020_3893940}. \textsc{tardis} operates by propagating photon packets through a specified ejecta, calculating their interactions with the material to synthesise a spectrum. The wavelengths of the initial packets are sampled from a blackbody distribution and released from an opaque inner boundary, which represents the photosphere in the homologously expanding plasma. As time progresses and the densities fall, this boundary in the expanding material will recede inwards to reveal slower moving ejecta. It is upon this fact that the method of abundance tomography is built; each successive spectrum will have a slightly lower photospheric velocity and  therefore, will be constraining the abundances of this newly revealed lower velocity matter.

For each \textsc{tardis} simulation there are five inputs: the luminosity (L), the time since explosion (t), the photospheric velocity (v$_{\text{ph}}$), the abundance profile (A), and the density profile ($\rho$). The abundance and density profiles are comprised of a number of discrete shells defined at different velocities.

Spectra synthesised by \textsc{tardis} cover a range of wavelength domains, and as such this creates issues with the validity of this photospheric approximation \citep{2014MNRAS.440..387K}. Photons of different frequencies probe different optical depths in a plasma and therefore, experience photospheres at different locations. The simplification of the system to a single fixed photosphere across the spectrum results in a flux excess seen in the redder wavelengths, however the absorption features still show through and can be used for line identification.

Another discussion point in terms of validity is the lack of treatment of radioactive decay, as the luminosity in a SN Ia is driven by the decay of $^{56}$Ni to $^{56}$Co to $^{56}$Fe. This energy injection is encoded in the blackbody emission from the photosphere, however, the energy from the decaying $^{56}$Ni in the ejecta above the photosphere is not incorporated. The negative impact of this approximation is insignificant whilst the bulk of this radioactive material lies below the photosphere and thus the code produces synthetic data in agreement with observations up until several days after peak for normal SNe Ia \citep{2014MNRAS.440..387K}. This is in line with \cite{ShenLTEnonLTE} who found agreement within 0.2~mag between local thermal equilibrium (LTE) and non-LTE radiative transfer calculations at maximum light in optical passbands for normal SNe Ia, with deviations appearing in the 15 days that followed peak.

As a fainter and faster evolving target, SN~2021rhu is likely to transition to a nebular phase earlier than its normal thermonuclear counterparts, and as such non-photospheric aspects could emerge earlier, further limiting the validity of \textsc{tardis} modelling in terms of phase. \cite{Iax_TARDIS2} used \textsc{tardis} to model three SNe Iax which also fall into this fast evolving regime, with their models showing close agreement up until a few days after peak. If, however, we were to conservatively suggest this validity to break down some five days before maximum light, we would be left with a six epoch spectral series that well constrains the majority of the material in the model. Dropping these final two epochs would not affect the conclusions drawn from the modelling. As such, we include the modelling of all 8 spectra up until peak in this work, however highlight that the final two epochs may suffer from these early onset non-photospheric effects.

\subsection{Density profiles}
\label{density_profiles}
\begin{figure}
 \includegraphics[width = \columnwidth]{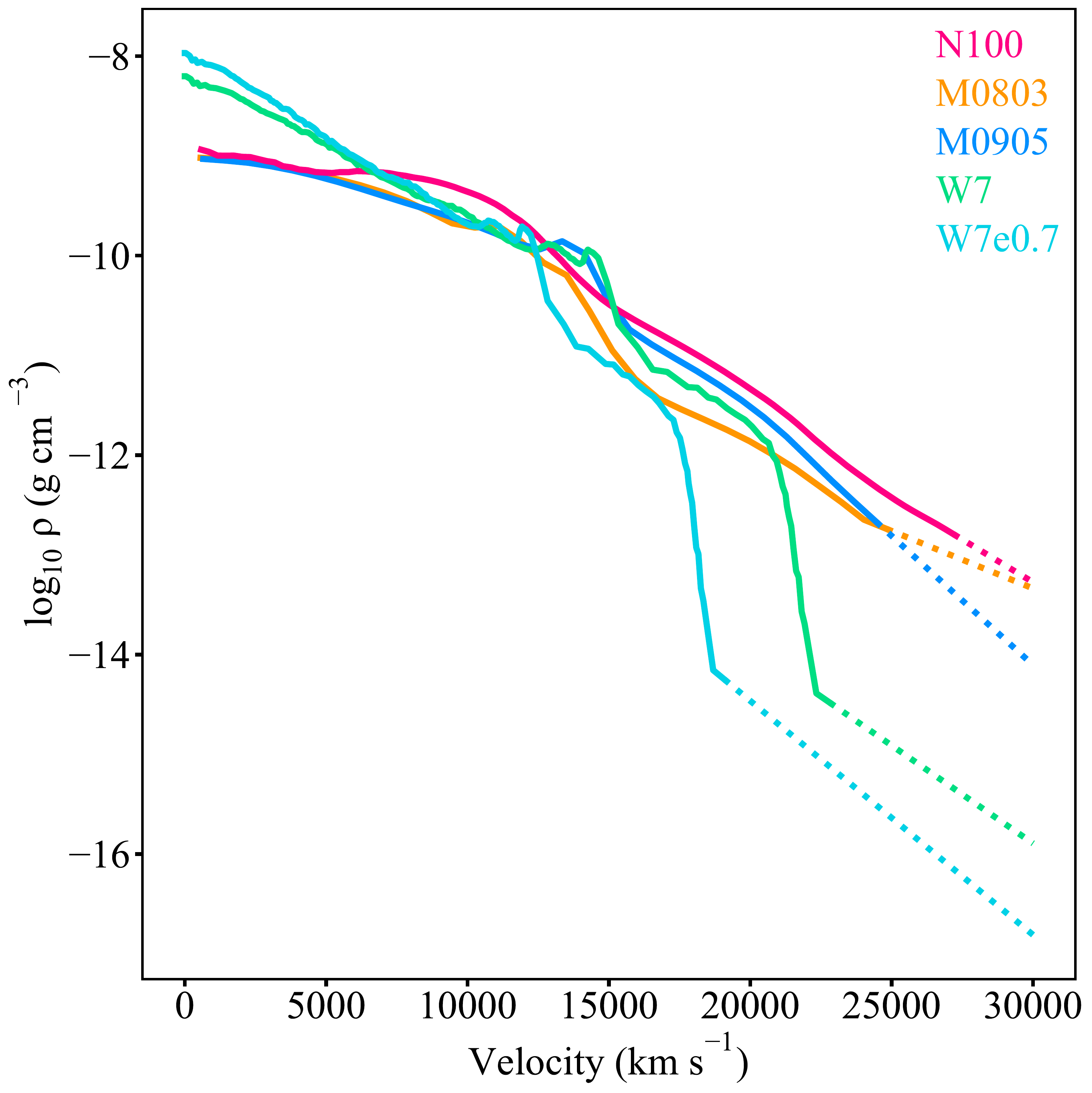}
 \caption{The density profiles of the five models discussed for SN~2021rhu expanded homologously to one day after explosion are shown as a function of velocity. These profiles arise from simulations of three different scenarios: double-detonation of a sub-Chandrasekhar mass WD (orange and dark blue), delayed-detonation of a Chandrasekhar mass WD (pink), and the fast deflagration of a Chandrasekhar mass WD (green and light blue). The light blue profile here is the scaled version of W7 chosen by \protect \cite{1986G_modelling} to best describe SN~1986G. The dashed regions at higher velocity are the linear extensions of the profiles in this logarithmic space with a single shell up at 30000~\kms. This extension was found to be necessary to reproduce the higher velocity components of the calcium features.}
 \label{fig:density_comp}
\end{figure}

In the context of SN~2021rhu, we investigated four different density profiles arising from three different explosion mechanisms (Fig.~\ref{fig:density_comp}). Firstly, we considered the N100 density profile resulting from the delayed-detonation of a Chandrasekhar-mass white dwarf \citep{N100, n100_2}. The delayed-detonation mechanism consists of an initial deflagration phase causing the white dwarf to swell before transitioning to a detonation. Secondly, we looked at two density profiles from double-detonation explosion models of sub-Chandrasekhar mass white dwarf, M0803 and M0905 from \cite{Sub_Ch_21rhu}. The M0905 profile was chosen as it has a comparable peak luminosity to SN~2021rhu, with the M0803 profile being added as a comparison model with different core and shell masses. These density profiles arise from hydrodynamical simulations of helium shells atop carbon-oxygen white dwarfs, assuming solar metallicity for the zero-age main sequence progenitors. The final density profile chosen was the W7 arising from the fast-deflagration of a Chandrasekhar-mass white dwarf \citep{W7_density}. The parameters of these models are given in Table \ref{tab:lit_models}. 

We also discuss the plausibility of W7e0.7, which is the W7 model scaled down to 70 per cent of the kinetic energy, as calculated in \cite{1986G_modelling} through the following equations,

\begin{equation}
\label{eqn:density_scaling_1}
\centering
    \rho' = \rho_0\left(\frac{E'}{E_0}\right)^{-\frac{3}{2}}\left(\frac{M'}{M_0}\right)^{\frac{5}{2}},
\end{equation}
\begin{equation}
\label{eqn:density_scaling_2}
\centering
    v' = v_0\left(\frac{E'}{E_0}\right)^{\frac{1}{2}}\left(\frac{M'}{M_0}\right)^{-\frac{1}{2}}.
\end{equation}
where $\rho$ is the density profile, $E$ is the kinetic energy, $v$ is the velocity profile, and $M$ is the mass. The initial values are denoted with subscript 0, with the new values indicated by dashes. The modelling work of SN~1986G by \cite{1986G_modelling} concluded this scaled W7 density profile to be the preferred choice, with the kinetic energy matching the explosion energy calculated from the derived abundance profile. They also found their sub-Chandrasekhar density profile to be insufficient as it required oxygen probing to layers in the ejecta deeper than sulphur, which is in direct conflict with nucleosynthetic calculations.

\begin{table}
\caption{Parameters for the four comparison literature models arising from the N100 delayed-detonation \citep{N100, n100_2}, W7 deflagration \citep{W7_density}, and the M0803 and M0905 double-detonation \citep{Sub_Ch_21rhu, doubledet_peaks} explosion models. The WD masses for the double-detonation models are given as core mass with the He-shell mass denoted in brackets.}
\label{tab:lit_models}
\begin{tabular}{ccccc}
\hline
\textbf{Model} & \textbf{WD Mass (M$_\odot$)} & \textbf{$^{56}$Ni Mass (M$_\odot$)} & \textbf{M$_{B,\text{peak}}$} \\ \hline
N100 $_\textit{a}$ & 1.406 & 0.604 & -19.0 \\
W7 $_\textit{b}$ & 1.378 & 0.587 & -19.1 \\
M0803 $_\textit{c}$ & 0.803 (0.028) & 0.130 & -17.0 \\
M0905 $_\textit{c}$ & 0.899 (0.053) & 0.382 & -18.3 \\ \hline
\end{tabular}
\end{table}

As the spectral modelling of SN~1986G only commenced 3~d before the $B$-band peak, the material at velocities above $\sim15000$~km~s$^{-1}$ is largely unconstrained by the modelling of \cite{1986G_modelling} . The earlier epochs captured by the spectral sequence of SN~2021rhu give us data with which we can constrain the faster moving ejecta and help distinguish between explosion models.  It was found that material at velocities higher than the maximum velocities in the density profiles was required to fill out the bluer end of some of the features in the spectra. To this end, each of the density profiles were extended with a single shell up at $30000$~km~s$^{-1}$. The density of this shell for each profile was calculated as a linear extrapolation in logarithmic space. These density profile extensions are shown as the dotted lines in Fig.~\ref{fig:density_comp}, and their importance will be discussed for each of the profiles in Section~\ref{results}.

\subsection{SN~2021rhu modelling with literature abundance profiles}
\label{sec:literature_models}
The initial abundance profiles used as input to the \textsc{tardis} models are the literature abundance profiles corresponding to the four density profiles in Section \ref{density_profiles}: N100, M0803, M0905, and W7. The luminosity $L$ was chosen so that the simulated spectra would align with the observations, the photospheric velocity $v_{\text{ph}}$ was chosen to match the \SiII\ velocity, and the time since explosion was set to agree  with the date of first light, with the inclusion of a 1~d dark phase to account for the offset between explosion and the first escape of photons \citep{Piro_darkphase}. A number of \textsc{tardis} simulations were then performed, varying these input values with the abundance and density profiles fixed, allowing us to assess the ability of these models to reproduce the observed spectral series. For each epoch we varied the photospheric velocity over a range of 3000 \kms and the time since explosion over a range of 2~d.

\begin{figure}
 \includegraphics[width = \columnwidth]{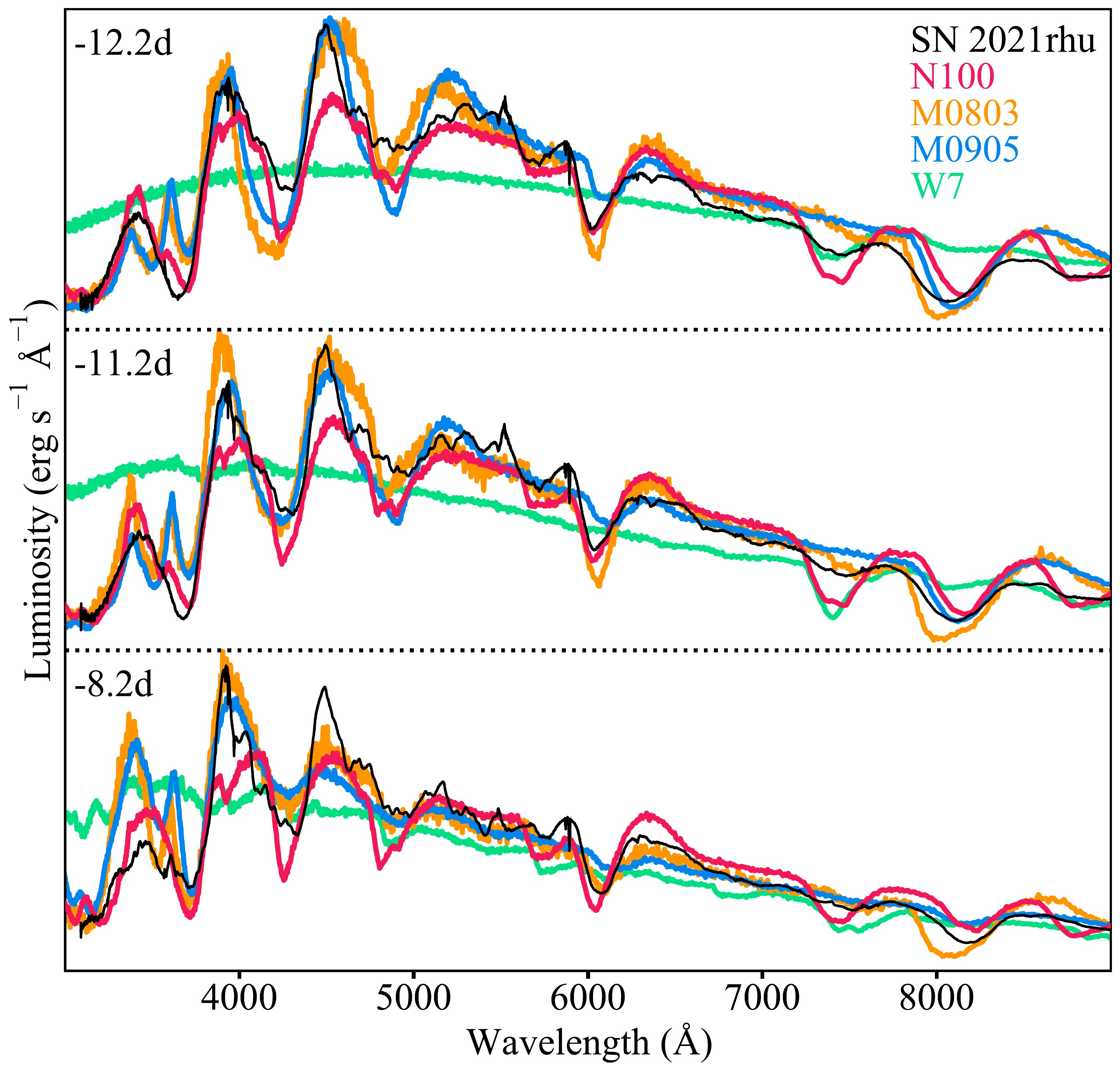}
 \caption{Best matching simulated spectra synthesised with the base literature density and abundance profiles from W7, M0803, M0905, and N100.}
 \label{fig:raw_models}
\end{figure}

Figure~\ref{fig:raw_models} shows the results of fitting these literature abundance and density profiles (`base' models) to the three highest resolution XShooter spectra at $-$12.2, $-$11.2, and $-$8.2~d with respect to maximum light. The first synthetic spectrum resulting from the fast deflagration W7 model is featureless with the exception of the \OI\ feature at $\sim7500$~Å. In the interest of matching observed line velocities at these early phases, $v_{\text{ph}}$ is required to lie above $\sim15000$~km~s$^{-1}$, and in the case of W7 this high-velocity region only contains oxygen, carbon, and neon. Some weak features begin to form in the following two epochs with the falling of the photospheric velocity, however the base W7 model is far from resembling the spectra of SN~2021rhu. The ejecta of the scaled W7e0.7 are even more compact than the W7 as visible from the density distribution in Fig.~\ref{fig:density_comp} and would as such be even less capable of matching the observations here.

The double-detonation M0803 and M0905 models reproduce the temperature and overall shape of the earliest spectrum fairly well (orange and blue lines in Figure~\ref{fig:raw_models}), with the temperature rising too quickly for the other two epochs resulting in a large flux excess in the blue. The absorption features are at roughly the correct velocities, however with strengths that do not resemble the features in the observed spectra. The signature \SiII\ $\lambda6355$ line, like the \CaII\ NIR, is too strong in the M0803 spectra, with this same silicon feature being too weak in those formed by M0905. Both of these models fail to reproduce the higher velocity component to the \CaII\ H\&K absorption structure. The key \TiII\ trough at $\sim4300$~Å is far too strong in the earliest epoch, owing to the large amounts of titanium that is synthesised in these double-detonation models, with this feature weakening in the later spectra due to the rising temperatures.

While the N100 literature model cannot reproduce the observed spectra entirely, it is the closest of the base models to doing so. With the overall spectral shape replicated for all three epochs, the temperature evolution resembles SN~2021rhu. However, like the double-detonation models, the feature strengths are a source of discrepancy in this model. Similarly to the double-detonation models, the delayed-detonation N100 is incapable of filling out the higher velocity end of the \CaII\ H\&K absorption complex. There is also an over-absorption in the region of the \TiII\ feature but in the case of N100, this stems from the large amounts of magnesium, which also produces an absorption line in this region. Finally, the N100 has an overabundance of Si when compared to SN~2021rhu. This can be seen not only in the excess absorption at $\sim4900$~Å, but also in the gradually growing P-Cygni emission from the \SiII\ $\lambda6355$ feature across the three epochs. The velocity of this synthesised silicon feature shows very little evolution across the spectra, despite the decreasing $v_{\text{ph}}$, once again pointing towards a larger than required silicon abundance above the photosphere.

In summary, while the delayed-detonation (N100) and double-detonation models (M0803 and M0905) provide reasonably similar spectra to those observed in SN~2021rhu pre-maximum, the lower levels of high-velocity material in the models and line strength discrepancies - in particular the mismatch in the \TiII\ absorption - suggest that using custom abundance profiles may provide closer matches to the data.

\subsection{Custom abundance profiles for SN~2021rhu}
\label{sec:custom_abund}
As evident from Fig.~\ref{fig:raw_models}, none of the tested base literature models are capable of faithfully replicating the spectral evolution we observe for SN~2021rhu. As such we require custom abundance profiles, derived through abundance tomography \citep{abundance_tomography}. This technique is used to develop a custom abundance profile for an assumed density profile and has been previously used to gain insights into the structure and composition of the ejected material of thermonuclear transients \citep{abund_strat2,abund_strat3,abund_strat4,Iax_tardis,abund_strat6}. Firstly, the abundances of the higher velocity material are constrained manually to match the features and shape of the earliest spectrum. Each successive spectrum has a slightly lower photosphere and therefore constrains a new lower velocity region of the ejecta. This is performed for as many of the spectra possible until the validity of the approximations made by \textsc{tardis} make it unfeasible to continue. Firstly we derived a custom abundance profile using the N100 density profile as the N100 model was the closest to matching the data (Fig.~\ref{fig:raw_models}). We note that as we are building a completely new abundance profile, the resulting model is no longer representative of the N100 explosion model or the delayed-detonation mechanism, it merely shares a density profile. Once we had a working model for the N100 density profile, we then used a `projection' or scaling method (described in Section \ref{sec:profile_projection}) to obtain models based on the other density profiles. The finalised models will provide insight as to the composition and structure of SN~2021rhu which will then be related back to signatures from the explosion models in Section~
\ref{sec:comp_to_lit_models}.

To distinguish the derived custom abundance profiles with the literature density profiles from the full models from the literature we use a * notation, e.g.~the original delayed-detonation N100  model from the literature is referred to as N100, whereas the model with the custom abundance profile that borrows the N100 density profile is denoted N100*. This same convention is used for W7*, M0803* and M0905*. 

\begin{table*}
\caption{\textsc{tardis} simulation parameters for the N100* and N100*host models. The * symbols denote that while the density profiles correspond to the literature N100 model, the abundance profiles are custom built for SN~2021rhu. N100*host is the model designed for the host extinction corrected spectra (see Section~\ref{sec:N100*h}).}
\label{tab:tardis_params_N100_W7}
\begin{tabular}{ccccccccc}
\hline
\multirow{2}{*}{\textbf{Phase}} & \multicolumn{2}{c}{\textbf{t} (d)} & \multicolumn{2}{c}{\textbf{v}$_{\text{ph}}$ (\kms)} & \multicolumn{2}{c}{\textbf{L} (log L$_\odot$)} & \multicolumn{2}{c}{\textbf{T}$_{\text{BB}}$ (K)} \\
& N100* & N100*host & N100* & N100*host & N100* & N100*host & N100* & N100*host \\ \hline
$-$12.2 & 6.0 & 6.5 & 16000 & 16000 & 8.38 & 8.56 & 7378 & 7788 \\
$-$11.2 & 7.0 & 7.5 & 14500 & 14500 & 8.5 3 & 8.71 & 8103 & 8622 \\
$-$10.2 & 8.0 & 8.5 & 14000 & 14000 & 8.68 & 8.88 & 8413 & 9177 \\
$-$8.2 & 10.0 & 10.5 & 13500 & 13500 & 8.93 & 9.13 & 8712 & 9555 \\
$-$7.4 & 10.8 & 11.3 & 13200 & 13200 & 9.04 & 9.24 & 9058 & 9988 \\
$-$5.4 & 12.82 & 13.32 & 12700 & 12700 & 9.19 & 9.39 & 9216 & 10218 \\
$-$1.4 & 16.78 & 17.28 & 11500 & 11500 & 9.31 & 9.53 & 9193 & 10242 \\
+0.6 & 18.79 & 19.29 & 11300 & 11300 & 9.34 & 9.57 & 8823 & 9819 \\ \hline
\end{tabular}
\end{table*}

\subsubsection{Projection to other density profiles}
\label{sec:profile_projection}
It is clear from Fig.~\ref{fig:density_comp} that the N100 density profile is very similar to those of the double-detonation models, M0803 and M0905 in the regions covered by our \textsc{tardis} modelling ($>$11000~\kms). Therefore, instead of performing abundance tomography for each of the double-detonation models, we use the following scaling or `projection' of the N100* onto the base M0803 and M0905 density profiles to obtain M0803* and M0905* custom abundance profiles for each species, X, given by 
\begin{equation}
    A_{\text{X, p*}}=\frac{A_{\text{X, N100*}} \times \rho_{\text{N100}}}{\rho_\text{p}}
    \label{eqn:mangle}
\end{equation}
where $A_{\text{X, p*}}$ refers to the new abundance profile of species X for the custom model p*, and $\rho_\text{p}$ is the density profile from the literature model p. For example, the scaled abundance profile for Si for the M0803* model ($A_{\text{Si,M0803*}}$) is calculated as the product of the abundance profile of Si in N100* ($A_{\text{Si,N100*}}$) and the ratio of the density profiles of the N100 to M0803 models ($\rho_{N100}/\rho_{M0803}$).  In each of these adjusted models, carbon is then used as a `filler species' to normalise the total mass fraction. The spectral evolution can then once again be calculated with \textsc{tardis} for these new models, retaining all the other input parameters from the N100* model.

In cases where the projection of the different species causes their abundances to sum to greater than 1, the filler carbon mass fraction is set to 0 and the species mass fractions are scaled down while retaining their relative proportions constant. This was not found to cause issues with the resulting synthetic spectra.

Compared to the density profiles from the double-detonation models, the W7 and the lower energy W7e0.7 model density profiles differ more significantly from that of the N100 in the region above 22000~\kms for W7 and 17000~\kms for W7e0.7. These large differences in the outer regions will be reflected in the synthetic spectra generated by the projected model W7*, namely in its ability to reproduce the high velocity features.

This method allows us to project between similar density profiles as the variations in ejecta conditions are minimal. This can in fact be expanded to more delayed-detonation \citep{n100_2}  and double-detonation models \citep{Sub_Ch_21rhu}, for which the scatter between the density profiles of models arising from the same explosion mechanism is similar to the scatter between the two mechanisms, making the density profiles degenerate.

\subsubsection{Modelling the host extinction corrected spectra}
\label{sec:ion_projection}
As detailed in Section~\ref{sec:observations}, the host component to the extinction correction was not included for the formation of the N100* model. Here we have recalibrated the observed spectra to account for this host extinction of $A_\text{V}$ = $0.42$ mag. In terms of changes to the \textsc{tardis} input parameters, the main difference will be an increase in the required luminosity as each of the observed spectra are now brighter. Similar to the projection of the N100* model to the other density profiles, we perform a similar type of projection to create an initial pass model for the reproduction of the host extinction corrected spectral series.

When we increase the luminosity input for a synthesised spectrum, this principally causes an increase in the temperature of the plasma. This in turn affects the ionisation balances of each of the species and as a result, reshapes the features in the final spectrum. In the spectra of SNe Ia, the largest contributors to the spectral features are the singly ionised species; being responsible for key structures such as the signature \SiII $\lambda6355$, the `w' shaped \SII, \CaII\ H\&K, and of course the \TiII\ trough. A similar effect accompanies the shifting of the time since explosion, with earlier times making for hotter ejecta and higher ionisation states. To this end we perform a projection not to simply retain the overall species densities, but specifically the singly ionised species densities. This can be achieved by identifying the required luminosity and time at each of the epochs to match the brightness and temperature of the host extincted spectra, and then running these simulations with the raw N100* model to extract the relative fraction of singly ionised material for each species. The projection for each species X is then calculated between  the initial luminosity L$_0$ and L with singly ionised mass fractions f using the following

\begin{equation}
    A_{\text{X, L, t}}=\frac{A_{\text{X, L}_0\text{, t}_0} \times f_{\text{X}^+\text{, L}_0\text{, t}_0}}{f_{\text{X}^+\text{, L, t}}}
    \label{eqn:ion_projection}
\end{equation}

Using Si again as an example, the resulting Si abundance profile at the new luminosity L, and time t ($A_{\text{Si, L, t}}$), is the product of the Si abundance profile with the original parameters ($A_{\text{Si, L}_0\text{, t}_0}$) and the ratio of the relative fraction of \SiII\ with the original parameters ($f_{\text{Si}^+\text{, L}_0\text{, t}_0}$) with the fraction of \SiII\ with the new parameters ($f_{\text{Si}^+\text{, L, t}}$). We refer to this model for the host extinction corrected observations as N100*host. The resulting synthesised spectra and analysis of this model can be seen in Section.~\ref{sec:N100*h}.

\begin{figure}
 \includegraphics[width = \linewidth]{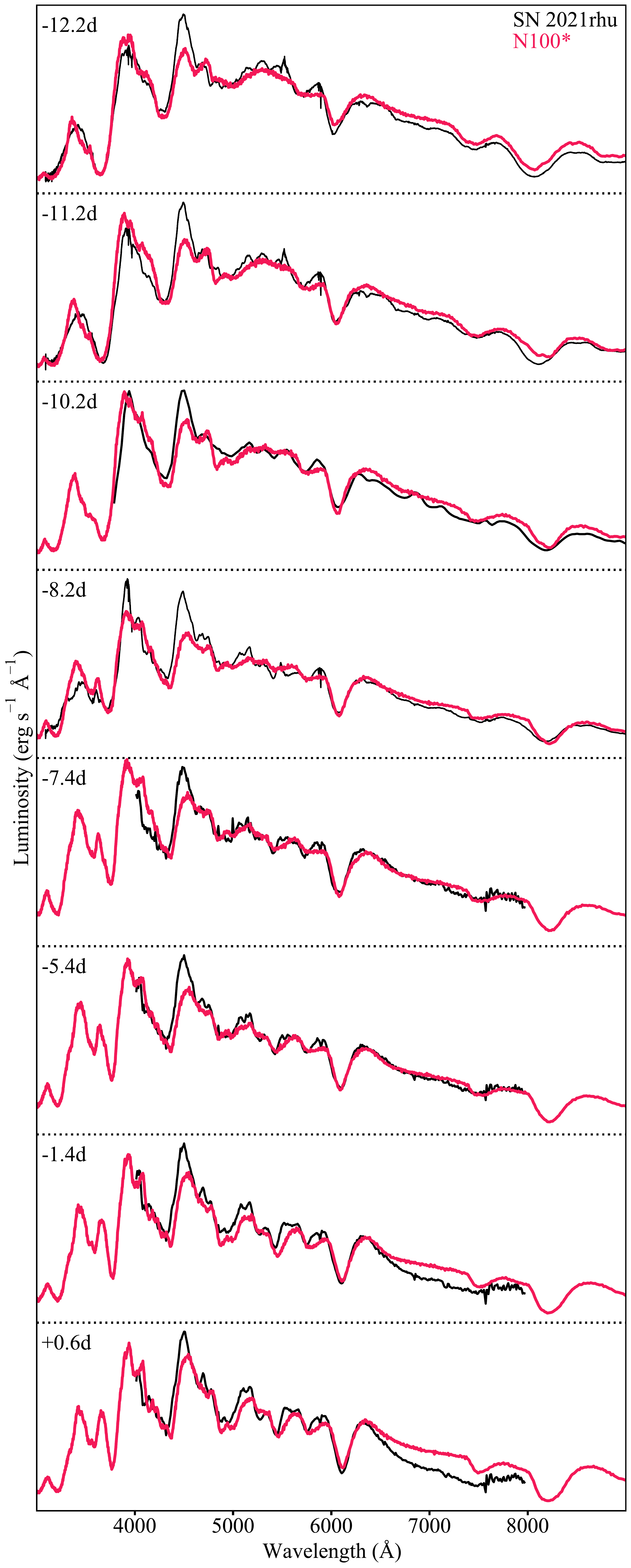}
 \caption{\textsc{tardis} simulations for the 8 spectra up to peak light with the fixed N100 density profile and custom abundance profile derived manually through abundance tomography (N100*). N100* is displayed in pink overplotted upon the photometrically calibrated spectra without the host extinction correction in black.
 }
 \label{fig:simulations_ddt}
\end{figure}

\section{Results}
\label{results}
In this section, we discuss the results of our modelling efforts and the link to the explosion models. A key feature of the spectra of transitional events like SN~1986G and SN~2021rhu is the strong \TiII\ absorption at $\sim$4300~Å. As discussed earlier, it is an open question as to whether the presence of \TiII\ absorption in these SNe is due to an increase in Ti abundance (that can be linked to explosion mechanism) or it is a temperature effect, where lower temperatures in these sub-luminous events produce conditions where \TiII\ absorption is more pronounced. In Section~\ref{sec:N100*} we detail the fiducial custom model N100* with the N100 delayed-detonation density profile. In Section~\ref{sec:W7*} we describe the results of the modelling using the W7* model, and in Section~\ref{sec:M0803*M0905*} we highlight the results of the two custom abundance profiles with the double-detonation density profiles. Finally, we present the results of the N100*host model for the host extinction corrected spectra in Section~\ref{sec:N100*h}.

\subsection{N100* - delayed-detonation density profile}
\label{sec:N100*}
The delayed-detonation N100 model is characterized by the highest densities in the outer layers of any of the models considered here (see Fig.~\ref{fig:density_comp}). The abundances of the original N100 model as function of velocity were presented in \cite{n100_2}. As discussed in Section \ref{sec:custom_abund}, we have derived new fully custom abundances using the abundance tomography technique (e.g.~sequential fitting of multiple epochs) with the N100 density profile to produce a best-matching model  (N100*) to the data. This density profile was chosen as the initial starting point as the raw literature N100 model provided the best match to SN~2021rhu in Fig.~\ref{fig:raw_models}. Once again we highlight here that the custom model N100* shares the density profile from the N100 explosion model, however, is independent of this explosion model due to the fully custom abundance profile.

\begin{figure}
 \includegraphics[width = \linewidth]{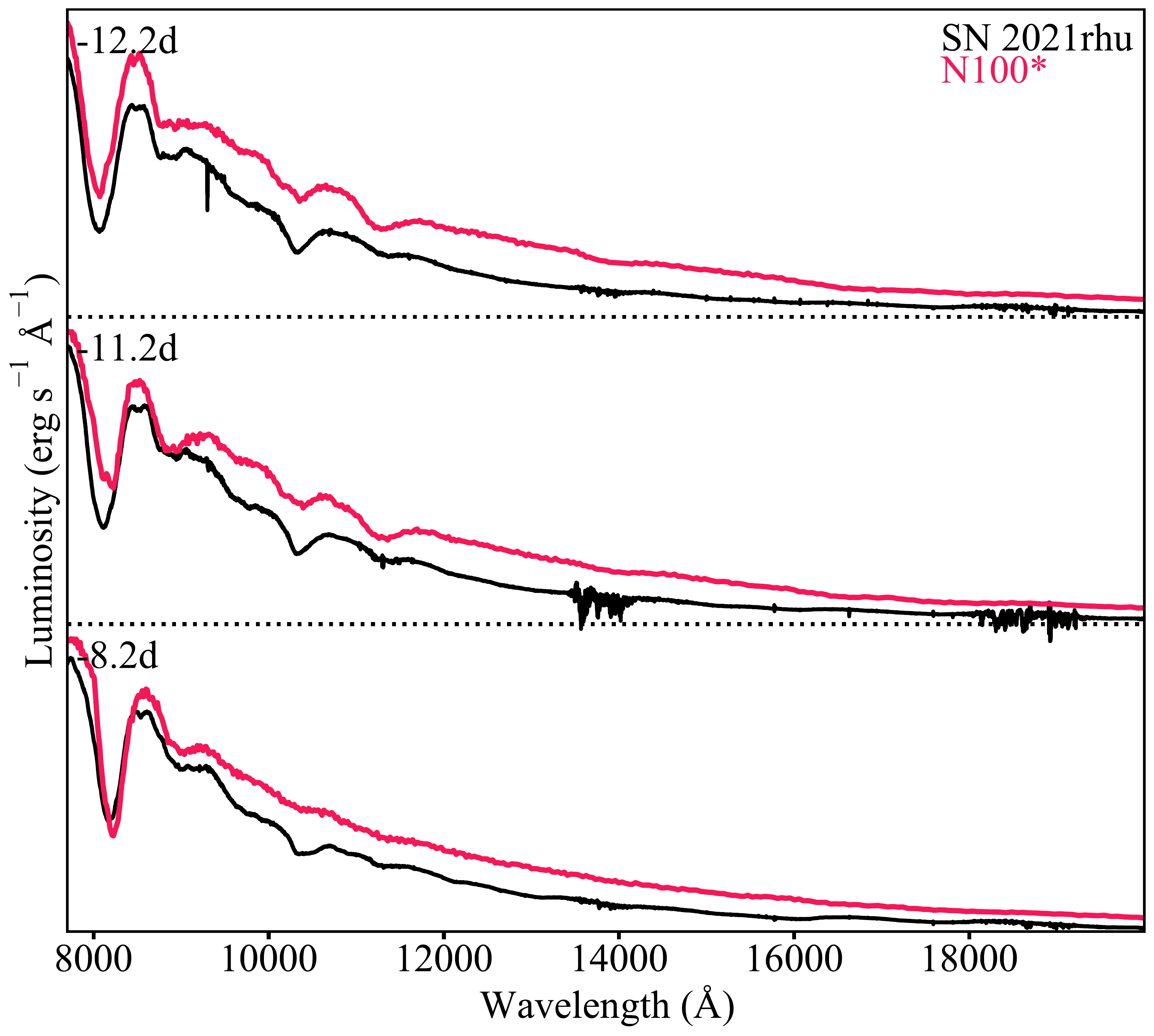}
 \caption{\textsc{tardis} simulations for the 3 XShooter epochs in the near-infrared using a fixed N100 density profile and custom abundance profile derived manually through abundance tomography (N100*). N100* is displayed in pink overplotted upon the photometrically calibrated spectra in black.
 }
 \label{fig:simulations_ddt_nir}
\end{figure}

With the fixed N100 density profile, the requested luminosity tuned to the brightness of each spectrum, and $v_{\text{ph}}$ roughly following the velocity of the \SiII $\lambda6355$ line \citep{Tanaka_Si_photosphere}, the time since explosion is the only free parameter to tune the temperature to match the overall spectral shape. From the fitting of the light-curve rise (see Section~\ref{sec:LC_properties}) we know the first spectrum to be $5.03\pm0.58$~d after first light (in the ZTF\textit{r}-band). Converting this to time since explosion then requires the inclusion of some dark phase to allow for the photons to escape the ejecta. As mentioned in Section \ref{sec:LC_properties}, the power-law index of this fit can be taken as a probe for the dark phase, and with an index close to the average value we can assume a dark phase of up to a couple of days. The best matching spectral shape was found to occur with the first spectrum of SN~2021rhu assumed to be at 6~d past explosion, which implies a dark phase of $0.97\pm0.58$~d.  This time since explosion is therefore in agreement with the fit to the early light-curve rise. The synthesised spectra with the N100* model can be seen in Fig.~\ref{fig:simulations_ddt} for optical wavelengths and in Fig.~\ref{fig:simulations_ddt_nir} for the three epochs of near-infrared coverage.  The best-matching \textsc{tardis} parameters for the models for each spectral epoch are listed in Table \ref{tab:tardis_params_N100_W7}.

\subsubsection{The earliest spectral modelling phase, $-$12.2~d}
The top observed spectrum of Fig.~\ref{fig:simulations_ddt} was recorded 12.2~d before maximum light with the XShooter spectrograph. Oxygen was initially used as a filler species to fill out the remaining mass fraction in the model. However, this caused a strong over-absorption of the \OI\ feature at $\sim$7500~Å, to the blue of the \CaII\ NIR. It was found that in order to reproduce the strength of this feature accurately we needed as little as a 3 per cent oxygen mass fraction in the outer ejecta. We instead turned to carbon as the filler species. While this allows the model to reproduce all the feature strengths accurately, this now leaves the N100* model with a mass fraction of carbon in these outer regions of 90 per cent. This amount of unburnt material is unrealistically high and will be addressed in Section~\ref{sec:C_reduction}.

\begin{figure}
 \includegraphics[width = \columnwidth]{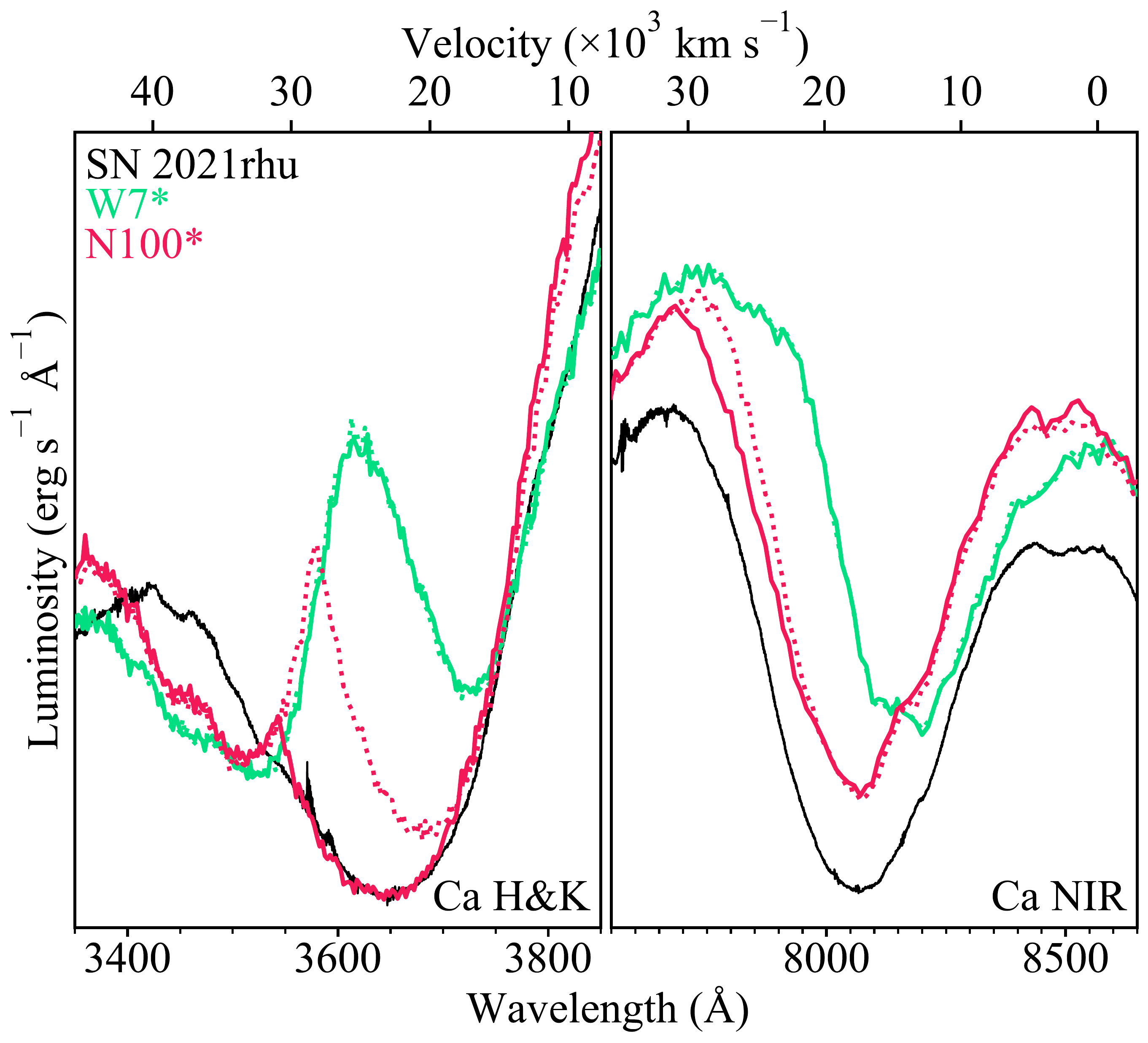}
 \caption{Zoomed view of the Ca\ H\&K and NIR features comparing the observed $-$12.2~d day spectrum of SN~2021rhu (black) against the synthesised spectra from N100* (pink) and W7* (green) with (solid) and without (dotted) the density profile extension up to $30000$~\kms. The extended shell has a calcium mass fraction of 1 per cent in the case of N100* and 100 per cent for W7* . The W7* model is discussed in Section~\ref{sec:W7*}}
 \label{fig:extension_Ca_comp}
\end{figure}

\begin{figure}
 \includegraphics[width = \columnwidth]{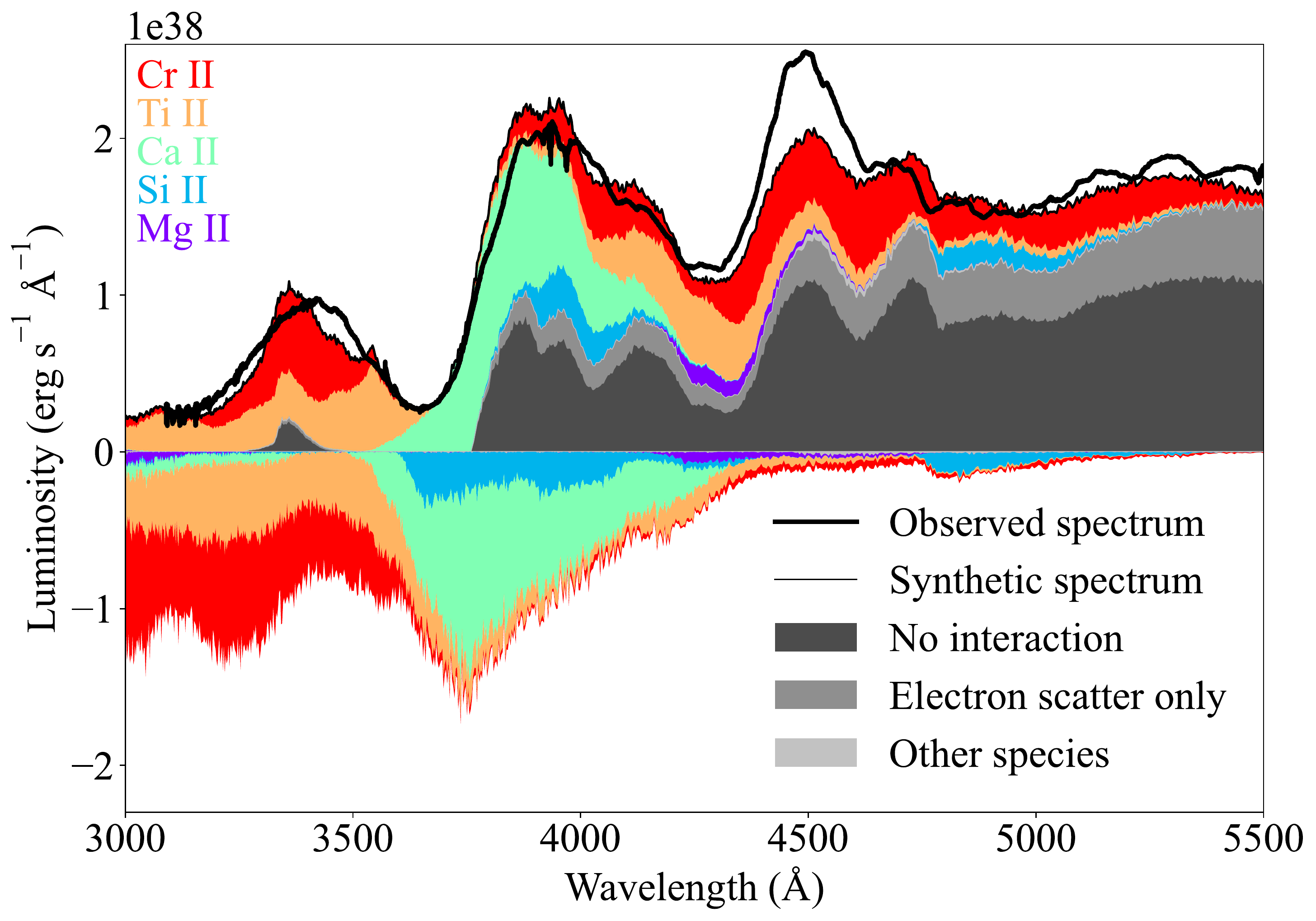}
 \caption{Spectral energy decomposition plot of the $-$12.2~d synthetic spectrum from the N100* model showing the contributions of different species to the signature \TiII\ feature. Shaded regions above zero flux represent emission with absorption shown below. The synthetic spectrum (thin black line), and the observed spectrum (thick black line) are also overplotted.}
 \label{fig:sdec_n100}
\end{figure}

In order to fill out the full velocity extent of the \CaII\ H\&K feature, it was necessary to extend the N100 density profile upwards with an additional shell at $30000$~\kms (see Fig.~\ref{fig:density_comp}). The calcium in this shell made up 0.05 per cent by mass with $\sim$96 per cent carbon and $\sim$3 per cent oxygen. A comparison of the synthesised spectrum, at this earliest epoch, with and without this density extension can be seen in Fig.~\ref{fig:extension_Ca_comp}. The shoulder contamination in the blue of the \CaII\ H\&K is produced by titanium and chromium and does not appear in the observed spectra until the $-$8.2~d spectrum. The wavelength range of the $-$10.2~d spectrum does not cover this feature and it is therefore unclear if this neighbouring feature is present at this epoch. Reducing the relative abundances of titanium and chromium can remove this feature although at great cost to the reproduction of the key feature of interest at $\sim$4300~Å.

As evident from the spectral energy decomposition of the synthetic $-$12.2~d spectrum in Fig.~\ref{fig:sdec_n100} in the 3000 - 5500~Å region, the so-labelled \TiII\ feature is in fact formed by a delicate balance of a few different species. While magnesium does typically show fairly strong absorption in this region, it was found that the evolution of the feature could be closely replicated with trace amounts of magnesium present in N100*. The peak at $\sim4500$~Å is under luminous in the synthetic spectrum compared to the observed data, a discrepancy which in fact remains throughout the epochs modelled. The largest contributor to this peak is seen to be chromium, meaning that an increase in chromium abundance could boost this flux. It is clear however that the chromium abundance greatly affects the spectral shape in the surrounding regions which match the observed data closely, and therefore, is well constrained.

Fig.~\ref{fig:simulations_ddt_nir} shows a comparison between the synthesised spectra and the XShooter epochs in the NIR region. There is an overall flux excess seen in the simulated spectra occurring as a consequence of the photometric approximation in \textsc{tardis}. The features however show through with their strengths and velocities ideally matching the observed absorption lines. The strongest feature in the blue end of this region is the \CaII\ NIR, with the two features to the red at $\sim8900$Å and $\sim10200$Å being predominantly formed by a blend of magnesium and carbon. As mentioned before the N100* possesses unrealistically large amounts of carbon in the outer regions and as such these two features would be far weaker in the synthetic spectra without this additional carbon.

\subsubsection{Pre-maximum spectral epochs $-$11.2 to $-$5.4~d}
The second XShooter spectrum was observed just a day after the first, at a phase of $-$11.2~d with respect to maximum light. The shape of the spectrum is very similar to the first epoch but has a higher luminosity and slightly slower velocities. Similar to the first synthesised spectrum, the $-$11.2~d model spectrum exhibits a blue shoulder to the \CaII\ H\&K which is absent in the observed spectrum, as well as an underproduction of the luminosity of the $\sim4500$~Å peak. Between the first and second epoch we require a large increase in the titanium mass fraction from 0.002 to over 1 per cent along with a slight boost in chromium from 0.01 to $\sim$0.04 per cent. It was also found that the carbon mass had to be dropped completely between the first and second photosphere to avoid forming the \CII\ 6580~Å line in the later spectra towards peak. This was instead replaced with oxygen. Due to 99 per cent of the oxygen in this region being mostly in the singly ionised state at this epoch, this does not affect the strength of the well matched \OI\ feature around 7500~Å. This ionisation balance will gradually sway towards neutral oxygen with successive epochs and causes a boost to this feature in the final two spectra ($-$1.4~d and $+$0.6~d). This reduction of carbon does not visibly reduce the contribution to the \CI\ feature just above $10000$~Å in the NIR spectrum at this epoch.

The $-$10.2~d spectrum was obtained with the P60+SEDM, which has a much lower spectral resolution and smaller wavelength range than XShooter and it does not cover the the \CaII\ H\&K feature. Therefore, it is unclear at which point between $-$11.2~d and $-$8.2~d the double peaked nature of this feature appears. At this epoch the best-matching model spectrum exhibits excess \SiII\ absorption at $\sim5000$~Å. This is accompanied by the enhanced absorption in the principal \SiII\ $\lambda6355$ feature. With the photospheres of the second and third epochs only separated by $500$~\kms we are fairly limited in our freedom to alter the silicon abundance.

The $-$8.2~d spectrum is the first point in which we stray from a one-day cadence, and also the final epoch for which we have an XShooter spectrum. The $v_{\text{ph}}$ has dropped by 500~\kms,  with an increased luminosity and temperature compared to the previous epoch. The shoulder feature to the \CaII\ H\&K starts to show through in the observed spectra at this point. While the model matches the velocities very well, as well as the \CaII\ absorption, there is a slight boost to the flux in the Ti/Cr shoulder along with the neighbouring peak to the blue. This is likely to be the result of a temperature that is too high, however raising the photosphere to reduce the temperature would disrupt the other well matching features. As discussed in Section~\ref{TARDIS}, the photospheric approximation made by \textsc{tardis} can cause such issues as a single photospheric velocity is assumed across the spectrum instead of it varying with wavelength; this therefore may be the cause behind this slight boost in flux. With the carbon drop-off introduced above the $-$11.2~d photosphere, the contribution of carbon to the magnesium feature in the NIR has faded completely, leaving a trace feature produced by the remaining magnesium in the model. This is no longer in good agreement with the data in which there still exists a clear feature.

From the $-$7.4~d spectrum onwards we have only spectra coming from the SPRAT instrument on the LT, which while retaining a fairly high spectral resolution brings a significant reduction in wavelength coverage (rest frame $\sim 4000-8000$~Å). Therefore, the spectra miss the \CaII\ H\&K and the neighbouring peaks in the blue end, and the \CaII\ NIR triplet in the red. The remaining features in between however are shown to be reproduced well in the remaining few simulations. As previously seen, the photospheric velocity continues to decrease while the luminosity and temperature increase for both these spectra. For an accurate reproduction of the spectral shape for the remaining epochs - most importantly for the $-$7.4~d and $-$5.4~d spectra - iron is required in the ejecta. In the N100* model the iron in the ejected material extends out to $13000$~\kms, making up 1 per cent of the mass in each of those shells.

\subsubsection{Maximum light spectra and comparison to SN~1986G modelling}
The $-$1.4~d epoch is the first epoch for which a direct comparison can be made between our N100* model and the model produced by \cite{1986G_modelling} for SN~1986G. The N100* spectrum was synthesised with $v_{\text{ph}}=11500$~\kms, a luminosity of $7.81\times10^{42}$~erg~s$^{-1}$ and a converged blackbody temperature of $9193$ K. The preferred SN~1986G model with the reduced energy W7e0.7 density profile at the $-$1 d epoch required $v_{\text{ph}}=8800$~\kms, a luminosity of $3.98\times10^{42}$~erg~s$^{-1}$ and a blackbody temperature of $9500$ K. The large difference in photospheric velocity here is not reflective of the difference in the velocities of the \SiII\ $\lambda6355$ feature between the two objects at this phase. Retracting our photosphere in this far raises the temperatures to $\sim11000K$ and negatively impacts the fit. This large difference between the two is likely due to the large differences between their chosen density profile and the N100 profile explored here; far larger than any differences between the N100 profile and the density profiles coming from the M0803 and M0905 models. The lower luminosity is to be expected in \cite{1986G_modelling} as SN~1986G is less luminous than SN~2021rhu. Large increases in the nickel and iron abundances were required for the modelling of the final two spectra of SN~2021rhu, with the mass fractions increasing to 30 per cent for nickel and 1 per cent for iron. As trace amounts of iron group material was required in the ejected material above this $11500$~\kms photosphere, the need for a fairly compact nickel distribution is well constrained.

Once again, the +0.6~d epoch lines up nicely in time with a reference spectrum for SN~1986G taken at +1 d. The N100* photosphere is placed at $v_{\text{ph}}=11300$~\kms, with a luminosity of $8.37\times10^{42}$~erg~s$^{-1}$ and a converged blackbody temperature of $8823$ K, marking the first point in which the temperature of the blackbody has decreased in the synthetic spectral series. For the SN~1986G model, the input parameters were $v_{\text{ph}}=7800$~\kms and a luminosity of $4.37\times10^{42}$~erg~s$^{-1}$ making for a blackbody temperature of $9700$ K. 

Due to the large proportion of radioactive nickel in the ejecta, and the 5 day time jump that would be required, the modelling of the +5.6~d spectrum is unfeasible with \textsc{tardis} and as such we cease the photospheric modelling here. 

\subsubsection{Summary of delayed-detonation N100* comparison }
Overall, the N100* model recreates very well the observed spectral series of SN~2021rhu in terms of spectral shape, feature strengths, and line velocities, as is seen in the spectral sequence in Fig.~\ref{fig:simulations_ddt} and Fig.~\ref{fig:simulations_ddt_nir}. The \textsc{tardis} parameters used for each of these simulations as well as the resulting blackbody temperatures can be found in Table~\ref{tab:tardis_params_N100_W7}. There do exist a number of discrepancies between the synthesised spectra and the observations but they are still minor. The main differences include the consistent underproduction by the models of the $\sim$4500~Å peak throughout the evolution, and the contamination of titanium and chromium in the blue wing of the \CaII\ H\&K in the earliest two epochs. We were unable to produce the previously mentioned \CI\ $\lambda6580$ feature which appears to be present in the first four spectra. In order to faithfully recreate the rest of the features, the photospheres for each of these early simulations were required higher than the observed velocities of this carbon feature.

\subsubsection{Comparison of N100* to N100 literature abundances}
The custom N100* model differs from the literature N100 model in a number of ways. Firstly, the shell extension up to $30000$~\kms does not exist in the base model, with the calcium abundance in N100 tapering off below 0.001 per cent around $20000$~\kms. Secondly, while the N100* titanium and chromium follow the rough velocity distribution found in N100, we require significantly enhanced mass fractions of titanium in the region around $11000$-$15000$~\kms to produce the evolution of the titanium trough. The iron and nickel content of N100* is more compact than N100 as these elevated levels were only required for the final two spectral epochs. Increasing the iron and nickel mass fractions at higher velocities caused disagreement with the data. Finally, the extreme carbon abundance found in N100* through its use as the filler species indicates that we require much less material in these outer regions. This will further be explored in Section~\ref{sec:C_reduction}.

\subsection{W7* - deflagration density profile}
\label{sec:W7*}
Here we show the results of the projection of the N100* model to the W7 density profile to create W7*. It is noted here that while the abundance and density profiles change as a result of the projection, the luminosity, time since explosion, and photospheric velocity parameters are kept constant between N100* and W7*. In Fig.~\ref{fig:density_comp}, it can be seen that above the threshold of $\sim15000$~\kms explored in the modelling of SN~1986G, the deflagration density profile W7 is significantly lower than the delayed-detonation (N100) and double-detonation (M0803, M0905) profiles and shows a sharp drop off in density. Figure \ref{fig:extension_Ca_comp} shows the impact of these low densities in the outer ejecta (not probed by SN~1986G spectra) on the \CaII\ H\&K and \CaII\ NIR triplet features in the earliest $-$12.2~d spectrum of SN~2021rhu. 

The W7* model is found to be a bad match to the observed data in this region because it fails to match the features from the high-velocity material and as a result the raw W7 density profile is insufficient in the context of the early evolution. The lower energy W7e0.7 density profile is more compact than the raw W7 density profile and, therefore, has even lower densities in these outer regions. We, therefore, also rule out the favoured density profile (W7e0.7) for SN~1986G as a candidate model for SN~2021rhu. 

As discussed in Section~\ref{density_profiles}, we extended the density profile of the W7* model with a shell at $30000$~\kms to see if this provides a better match to the data. Even filling this extension shell to be 100 per cent Ca only strengthens the synthesised feature slightly and still greatly under-produces the \CaII\ H\&K and \CaII\ NIR triplet features when compared to the observed spectrum. From this we can also conclude that the extended W7 density profile is not feasible for this object.
When comparing to the literature W7 model, the only species present in the ejecta above $\sim14000$~\kms are oxygen, carbon, and neon, meaning that even if the density extension was sufficient for the higher velocity material, the required abundance profile would differ greatly from the deflagration prediction in the literature.
The resulting synthesised spectra for W7* can be seen in the left-hand panel of Fig.~\ref{fig:mangled_models}.

\begin{figure}
 \includegraphics[width = \linewidth]{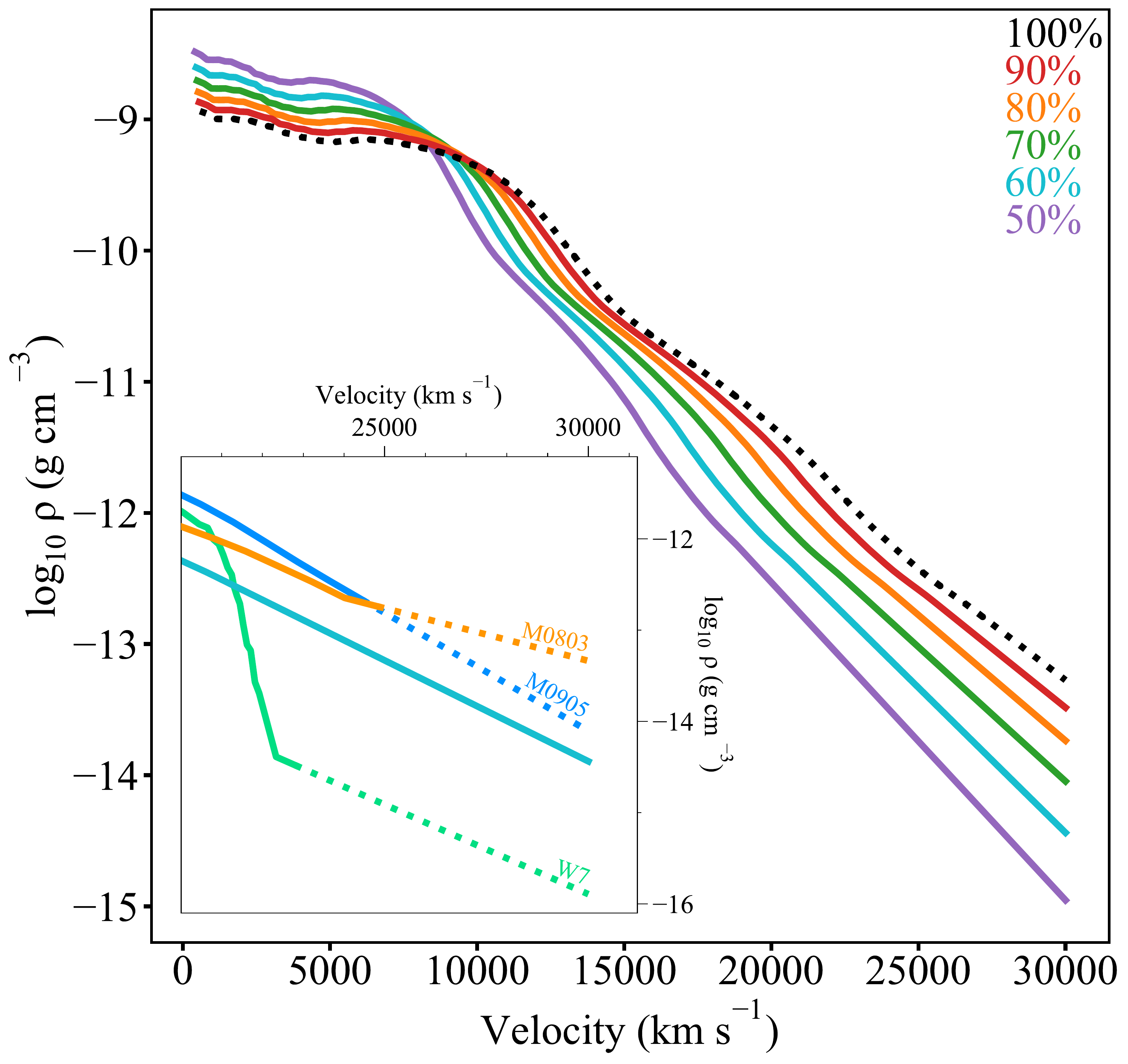}
 \caption{The result of the energy scaling upon the N100 density profile. The profiles with lower kinetic energy are less able to accelerate the ejected material to high velocities and therefore possess a more compact ejecta structure. The black dotted line represents the base N100 density profile with the 30000~\kms extension. The inset displays the location of the high velocity tail of the preferred 60 per cent scaled density profile between the W7 and double-detonation profiles.
 }
 \label{fig:ek_scaling}
\end{figure}

\begin{figure}
 \includegraphics[width = \linewidth]{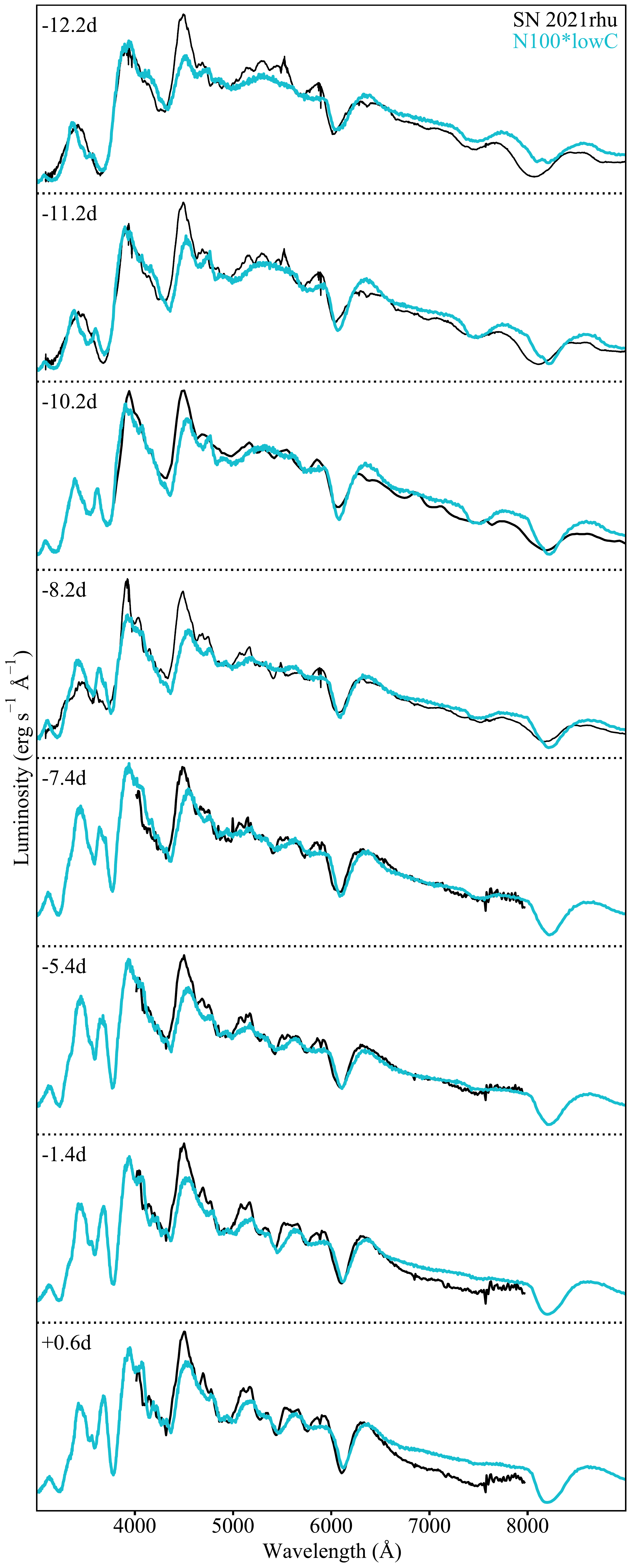}
 \caption{\textsc{tardis} simulations for the eight spectra up to peak light for the reduced carbon model N100*lowC calculated as the projection of N100* onto the 60 per cent kinetic energy scaled N100 density profile. N100*lowC is displayed in blue overplotted upon the photometrically calibrated spectra without the host extinction correction in black.
 }
 \label{fig:N100_60*}
\end{figure}

\subsection{M0803* \& M0905* - double-detonation density profiles}
\label{sec:M0803*M0905*}
In this section, we compare the observed spectra of SN~2021rhu to those of the projected models with the sub-Chandrasekhar double-detonation density profiles, M0803* and M0905*. As was the case for W7*, these models retain the luminosity, time since explosion, and photospheric velocity parameters used in N100*. We note that the density profiles used in this projection are those with the linear extrapolation in log space up to $30000$~\kms as discussed in Section \ref{density_profiles}. The synthesised spectra for these models can be found in Fig.~\ref{fig:mangled_models}. These synthetic spectra exhibit similar discrepancies to those seen for the spectra generated by N100*. The switch of density profile causes differences in the radiative transfer calculations which in turn causes the subtle differences between the spectra from these three models. As these projections are based upon the N100* model, these small differences mean that while they match the spectral evolution of SN~2021rhu well, they are not quite as close as N100*. The changes that would be required to the abundance profile to reconcile these differences are negligible and do not alter the conclusions or comparisons with literature models in Section~\ref{sec:comp_to_lit_models}.

When looking at the synthetic spectra from the M0905* model we see the formation of the \CI\ 6580 Å feature in the epochs from $-$8.2~d up until the final spectrum at peak. This appears as the M0905 density profile is greater than that of the N100 in a small region just below 15000 \kms. The normalisation of the abundance profile after projection causes the introduction of a carbon lump here which in turn is responsible for this feature showing through. This does not agree with the observed data and can also be seen in the synthetic spectra from W7* for the same reason. This carbon lump could be removed and filled in by the remaining species in the model without great negative impact to the match to the data.

When looking at the temperatures from the simulations with the different models and the same input parameters, they are found to be in good agreement with differences of $<$10 per cent between the most extreme models (W7* and N100*). Here we reiterate that it is not the temperature difference that rules out W7*, but the low densities in the high velocity regions being insufficient for the formation of high velocity Ca features.

\subsection{Reducing the carbon - N100*lowC}
\label{sec:C_reduction}
As outlined in Section~\ref{sec:N100*}, through the use of carbon as the filler species in the production of the N100* model, the resulting profile is comprised of $\sim90$ per cent carbon by mass fraction in the region above $\sim15000$~\kms. With the M0803 and M0905 density profiles being slightly smaller in this higher velocity regime than the N100, the projected models M0803* and M0905* contained slightly lower carbon abundances of 80-90 per cent.

We are able to constrain the inner velocity extent of the unburnt carbon ($>$15000 \kms), as such a high mass fraction deeper into the ejecta would produce discrepancies between the synthesised spectra and the observations towards peak. We are unable however to put solid constraint on the abundance of carbon in these outer regions and therefore choose to reduce the carbon to be equal to the oxygen abundance. Literature models such as the N100 and the W7 have carbon and oxygen mass fractions of the same order in the outermost regions. To this end, we turn to the kinetic energy scaling formulation in Equations~\ref{eqn:density_scaling_1} and \ref{eqn:density_scaling_2}. A lower kinetic energy version of the N100 density profile would have smaller densities in this higher velocity region and therefore the abundance profile would need less padding out from the filler carbon in this regime.

Figure~\ref{fig:ek_scaling} shows the resulting density profiles when scaling the kinetic energy to a range of different fractions. Reducing this energy to 60 per cent was found to be sufficient to bring the carbon mass fraction to similar magnitudes as seen for oxygen. The N100* model was projected onto this scaled density profile using the same method as used for creating the M0803*, M0905*, and W7* models. The inset of Fig.~\ref{fig:ek_scaling} displays the location of this preferred 60 per cent profile in the context of the other literature density profiles. Required to be lower in density than the M0803 and M0905 density profiles as the M0803* and M0905* also possess massive amounts of carbon, this profile is also required to possess higher densities than the W7 profile in this high velocity regime to be able to produce the high velocity edges of the calcium features.

This reduced carbon model with the scaled density profile is labelled as N100*lowC, and the resulting synthetic spectral series can be seen in Fig.~\ref{fig:N100_60*}. As seen before for N100*, M0803*, and M0905*, the N100*lowC model reproduces the spectral evolution of SN~2021rhu well. With a more plausible amount of unburnt carbon material, this is our preferred model for the observations of SN~2021rhu without host extinction corrections. The abundance profiles for each species in the N100*lowC model can be found in the top panel of Fig.~\ref{fig:abundances_lowC_host}.

\begin{figure}
 \includegraphics[width = \linewidth]{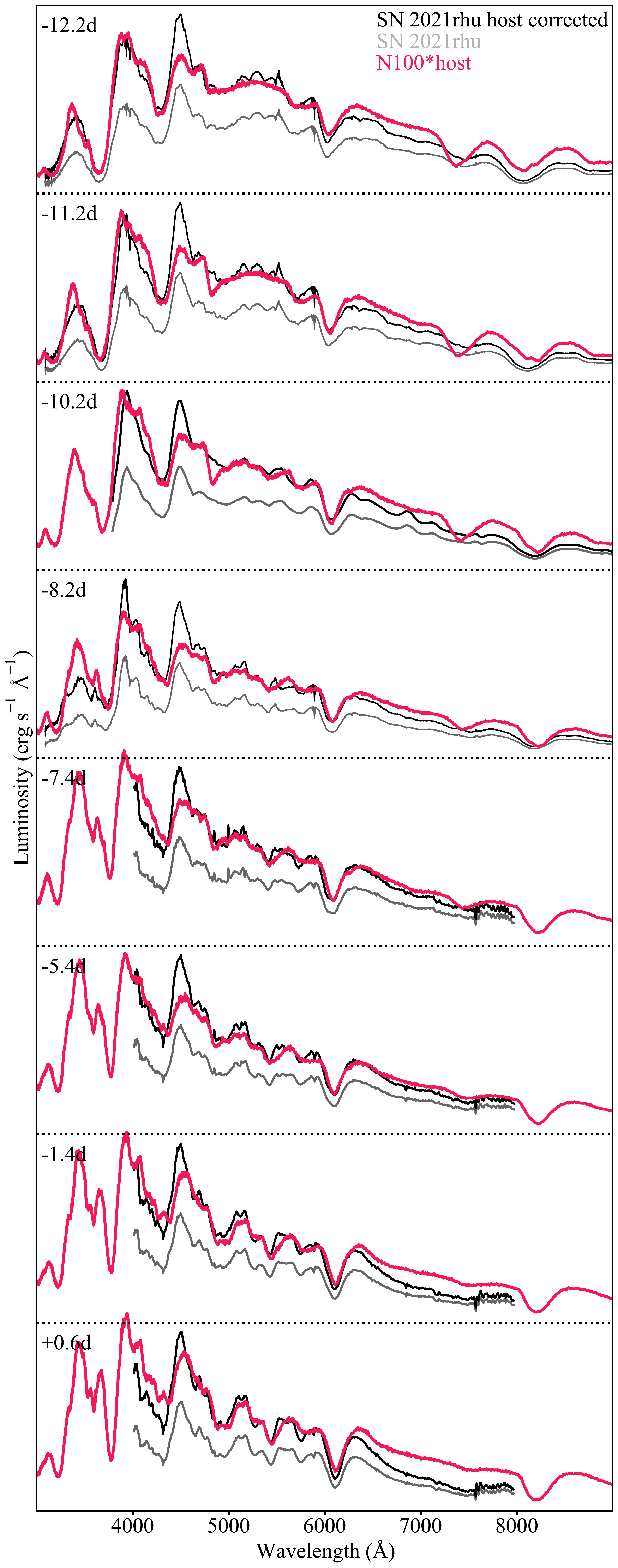}
 \caption{\textsc{tardis} simulations for the eight spectra up to peak light for the model to the host extinction corrected observations N100*host. N100*host is displayed in pink overplotted upon the photometrically calibrated spectra with the host extinction correction in black. The grey spectra are the observations seen for all the other models without the host extinction correction.}
 \label{fig:host_ext_sims}
\end{figure}

\begin{figure*}
 \includegraphics[width = \linewidth]{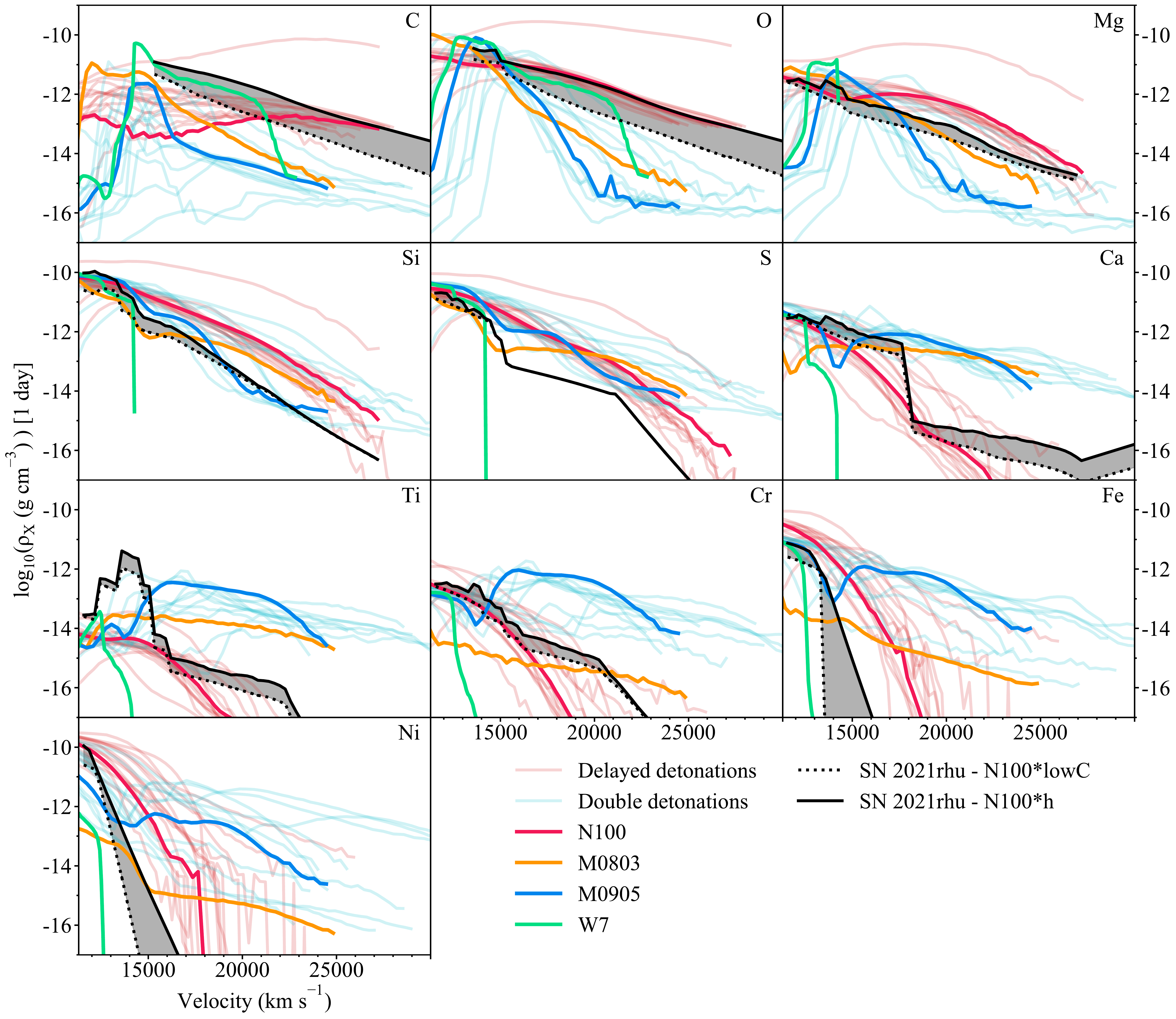}
 \caption{Comparisons of the derived upper and lower limits for the species densities of SN~2021rhu from the N100*host and N100*lowC model respectively, compared against the N100, W7, M0803, M0905 literature models. We stress here that the coloured lines are the raw literature models and not the custom models derived with the density profiles here. To outline the regions of the parameter space in which the different explosive mechanisms lie we also show the remaining delayed-detonation models from \protect\cite{n100_2} (faint red) as well as the other double-detonation models from \protect\cite{Sub_Ch_21rhu} (faint blue).}
 \label{fig:species_density}
\end{figure*}

\subsection{N100*host - corrected for host extinction}
\label{sec:N100*h}
A model to reproduce the higher luminosity observed spectral series in which we corrected for host extinction was produced through the projection method described in Section~\ref{sec:ion_projection}. The rising luminosities of the simulations bring with them rising temperatures and it is found that the 60 per cent kinetic energy scaling of the N100 density profile used for the N100*lowC model has densities which are too low in the region around $12000$~\kms to faithfully reproduce the later spectra. Therefore the projection is calculated from the base N100* model. Due to the sizeable increase of the IME abundances in the outer regions, the filler carbon content of the outer ejecta is greatly reduced from N100* to be in line with the oxygen abundance. This model for the host extinction corrected observations is labelled as N100*host. An increase in the time since explosion of 0.5 days was also found to be required to roughly match up the small peak at $\sim$3500~Å in the first two epochs.

As we constrain the amount of nickel in these abundance profiles not by the recreation of a certain feature, but by its effect on the plasma conditions and each spectrum as a whole, there is more room for movement in the its abundance profile. As we aim to produce an upper limit for each abundance with N100*host model we chose to inflate the nickel abundance as much as possible in this model until it began to compromise the match to the observed data in order to find this upper bound.

The synthesised spectra can be seen in Fig.~\ref{fig:host_ext_sims}. Once again this projection is able to well reproduce the observed data. As before, we argue that any modifications that would be required to tweak the model to match the data as well as N100* matches the observations without host extinction corrections, are insignificant for the comparison to literature models in Section~\ref{sec:comp_to_lit_models}.

This projection however does have two key drawbacks. Firstly it is built upon the assumption that the first ionisation state of each species is the only ion contributing to the final spectrum, the other ions are left unconstrained in this projection. While these are not the only contributing ions, they are the most important and are therefore required to be kept as constant as possible. The second drawback to this method is that the single ion mass fraction for the new luminosity L ($f_{\text{X}^+\text{, L, t}}$) is measured from the initial N100* model, and the very nature of changing the abundance balance will impact the state of the plasma and the ion balance of each species. This could be overcome by iterating the process two or three times to draw closer to keeping the ion density constant between N100* and the host extinction model, however we chose to make the small tweaks by hand. For this model carbon and oxygen were employed as normalisation species. The abundance profiles for each species in the N100*host model can be found in the bottom panel of Fig.~\ref{fig:abundances_lowC_host}.

\section{Discussion}
\label{discussion}
SN~2021rhu is a `transitional' SN Ia, with a luminosity between that of normal SNe Ia and the main class of underluminous SNe Ia, the 91bg-like SNe. It was discovered within a few days of first light in a very nearby (z=0.003506) spiral galaxy and its first spectrum was obtained within $\sim$5 days of first light. Along with its great dataset of early data, it is also of particular interest because of its use as a H$_0$ calibrator object \citep{Suhail_H0} and whether transitional events in general should be used in cosmological analysis. We have compared SN~2021rhu to a number of explosion model outputs using the \textsc{tardis} radiative transfer code to determine the most likely progenitor scenario and address these questions. 

\subsection{Building a preferred model for SN~2021rhu}
Our initial analysis involved comparing the observed spectra of SN~2021rhu to the literature density and abundance profiles for the Chandrasekhar-mass delayed-detonation model, N100 and the deflagration model (W7), as well as two sub-Chandrasekhar mass double-detonation models (M0803, M0905). The delayed-detonation (N100) and double-detonation models (M0803, M0905) have very similar density profiles (Fig.~\ref{fig:density_comp}) but their abundance profiles do differ, resulting in slightly different spectra (Fig.~\ref{fig:raw_models}). The N100 is found to match best to the data in terms of line strengths and velocities but the M0803 and M0905 are also to be in reasonable agreement. The W7 model does not match the early spectra of SN~2021rhu well due to the significantly lower densities in the outer ejecta and abundances in this high velocity region comprised solely of oxygen, carbon, and neon. A reduced kinetic energy model, W7e0.7, was preferred for another transitional event, SN 1986G \citep{1986G_modelling}, but this event lacked early spectra to test the significantly lower density ejecta predictions of the W7 and W7e0.7 models. 

Our next step was to develop preferred custom abundance profiles for SN~2021rhu using the abundance tomography technique to better match the spectral shape and features over the full range of spectra investigated here ($-12$ to +1 d from maximum light). This was achieved by firstly deriving a model based on the density profile of the closest matching literature model (N100), which we call N100*. This model was then scaled or `projected', as discussed in Section \ref{sec:custom_abund}, to produce two adapted sub-Chandrasekhar double-detonation (M0803*, M0905*) models and an adapted Chandrasekhar-mass (W7*) deflagration model. This projection is reasonable because of the general similarities of the density profiles of the N100 and double-detonation explosion models shown in Fig. \ref{fig:density_comp} and because the density profiles of the delayed-detonation and double-detonation models can provide a physically plausible description of the ejecta in SN~2021rhu.

Given the uncertainty in the amount of host galaxy extinction present, we also designed a model to match the host extinction-corrected spectra, which we call N100*host. We also determined a final preferred delayed-detonation model to have a more reasonable C to O ratio in the outer ejecta with a final preferred model called, N100*lowC using a simple scaling of the kinetic energy of the N100 density profile. Figure \ref{fig:ek_scaling} shows our preferred model without host extinction (N100*lowC) and Fig.~\ref{fig:host_ext_sims} shows our preferred model with host extinction (N100*host). The final double-detonation model spectra (M0803*, M0905*) are shown in Fig.~\ref{fig:mangled_models}, along with the W7* for completeness.

We find that the N100*lowC model matches the spectra of SN~2021rhu as well as the N100*, M0803*, and M0905* models, with the added advantage that it possesses realistic amounts of unburnt C in the ejecta. Across all models and all epochs, we consistently miss the full flux extent of the peak at $\sim4500$ Å. Another discrepancy that appears in all the custom models - however to varying degrees - is the two component nature of the absorption complex associated with the \CaII\ H\&K lines. While this dual component structure first shows up in the observed spectra at $-$8.2~d, excess absorption from titanium and chromium form this second component in the synthetic spectrum from as early as $-$12.2~d. Exclusive to the M0905* and W7* models, there is absorption from \CII\ $\lambda6580$ from $-$8.2~d onwards. This arises from excess filler carbon introduced by the density profile projection in regions where these density profiles (M0905 and W7) exceed the source density profile (N100). This is an artefact of the projection process and could be smoothed out by removing this excess carbon and normalising with the other species, with likely little effect on the rest of the spectrum.

\subsection{Delayed-detonation model preferred}
\label{sec:comp_to_lit_models}
As well as comparing the overall spectral feature matches between the observed SN~2021rhu spectra and the models in the spectral sequence, we have also investigated the comparisons on an element specific basis. One of the main conclusions from our modelling is that it is not the mass fraction model specific to the N100 density profile that matters, but the product of this with the N100 density profile itself. This product is seen as the numerator in Equation~\ref{eqn:mangle} and describes a specific density profile for each species (hereby referred to as a `species density profile`). As these species density profiles are independent of the underlying density profile, they can be compared against those from the explosion models in the literature. 

In Fig.~\ref{fig:species_density} we present comparisons of our preferred species density profiles for SN~2021rhu from the \textsc{tardis} modelling, with the species density profiles for the W7, N100, M0803, and M0905 literature models along with the rest of the double-detonation models from \cite{Sub_Ch_21rhu} and the rest of the delayed-detonation models from \cite{n100_2}. The solid and dotted black lines represent our preferred models with (N100*host) and without (N100*lowC) host galaxy extinction respectively. These models act as limits, with the species densities for SN~2021rhu lying somewhere in the shaded region between the two.\footnote{It is noted here that although the abundance and density profiles differ between N100*, M0803*, M0905*, W7* and N100*lowC, the species density profiles remain constant through the density profile projection. These four custom models are all therefore represented by the dotted lines in Fig.~\ref{fig:species_density} - with the exception of carbon panel as this is used as the normalisation species.} 

From inspection of the heavier species in the models in Fig.~\ref{fig:species_density}, it is evident that starting from calcium and moving to more massive elements, there exists a larger separation in this parameter space between the literature delayed-detonation of Chandrasekhar mass (e.g.~N100) and double-detonation sub-Chandrasekhar mass (e.g.~M0803, M0905) models. In each of these panels of heavier elements, the species density of SN~2021rhu is more consistent with the delayed-detonation models. Of the two main double-detonation models shown, SN~2021rhu is more consistent with M0803, which has a lower white dwarf core mass (0.8 \msun) with a smaller He-shell mass (0.03 \msun) than M0905, which has a core mass of 0.9 \msun\ and a He-shell mass of 0.05 \msun. However, both deviate more strongly from the SN~2021rhu range of values than the N100. 

There are two main differences between the N100 model and SN~2021rhu range in the heavier elements, i) a large bump in the titanium profile for SN~2021rhu at lower velocities and ii) more compact nickel and iron profiles are required for SN~2021rhu than seen in the NI100 model. The bump in the titanium profile was found to be necessary to consistently produce the trough feature in a number of spectra on the rise .This bump peaks to values more in line with many of the double-detonation models, however these models possess unreasonably high titanium abundances at higher velocities in the context of SN~2021rhu. We attempted to reduce this titanium peak by replacing it with magnesium or chromium as these species also produce absorption in this region however these were found to negatively impact the match to the data. Another option explored was to reduce the photospheric velocities in an attempt to flatten out this bunched up titanium however this in turn caused higher temperatures and different line velocities which again negatively impacted the match to the data. The more compact nickel and iron distributions were required so that the earlier spectra (produced a higher velocity) are not contaminated by lines of these elements. 

The carbon and oxygen profiles in Fig.~\ref{fig:species_density} are also found to be most similar to the N100 model. The magnesium profile for SN~2021rhu seems to lie somewhere between the two sets of literature models, and the silcon and sulphur profiles sit lower than the predictions from literature. None of the species from our models are consistent with the predictions from the literature W7 deflagration model. While unable to provide a strict answer to the question of the correct explosion mechanism, the ejecta composition and structure of our custom model for SN~2021rhu are more consistent with the Chandrasekhar delayed-detonation mechanism than the sub-Chandrasekhar double-detonation mechanism. However, more titanium is required than seen in the delayed-detonation N100 model to explain the later spectra when the ejecta is hotter and the strong titanium absorption features cannot be explained by simply a temperature effect.

\subsection{Is SN~2021rhu suitable for use in cosmological measurements?}
The properties of the transitional SNe Ia, such as SN 2021rhu, differ from `normal' SNe Ia in a number of ways, their light curves are generally less luminous at peak and they appear redder, due to the stronger absorption in the blue from \TiII. They also tend to have a stronger ratio of \SiII\ 5972 to 6355 \AA\ absorption, which is likely due to the cooler temperatures in these transitional events \citep{hachinger_siratio}. We have found that when comparing our custom model for SN~2021rhu to explosion models, the deflagrations and double-detonations are worse matches, with the custom composition lying roughly consistent with the delayed-detonation models. The double-detonation models are incompatible with SN~2021rhu, as seen in their species density profiles (see Section \ref{sec:comp_to_lit_models}) of calcium, titanium, chromium, iron and nickel because of their greater extent in velocity space that are intrinsic to the model. The species density profiles for SN~2021rhu are similar to those for the delayed-detonation models, with the two limitations being the bump in the titanium profile at lower velocities along with the more compact distributions of nickel and iron that still need to be explained. Therefore, overall this suggests that SN~2021rhu is more consistent with a lower luminosity (and lower temperature) version of a Chandrasekhar-mass explosion of a white dwarf undergoing a delayed-detonation. Although we note that full explosion simulations of a lower luminosity version of this model would need to be produced to confirm if this is possible.

The explosion scenarios of `normal' thermonuclear SNe remain undetermined. The double-detonation mechanism has been suggested to produce the range of normal SNe Ia \citep{shen_doubledet}. However, \cite{doubledet_peaks} found spectral simulations of double-detonation models to be too red at early times due to heavy line blanketing from the overabundance of heavier elements in the upper ejecta - this is in agreement with the findings here comparing SN~2021rhu to their models. Some delayed-detonation models have been found to agree with normal SNe Ia in terms of light curve and spectral evolution \citep{ddt_observables}, with the caveat of slightly redder colours around peak. From our spectral modelling, we have found SN~2021rhu to be most consistent with a low-luminosity delayed-detonation, therefore, we speculate that it could be part of a continuum of objects stretching out from the cosmologically useful SNe Ia. There is however a sizeable scatter around the Phillips relation from the delayed-detonation models from \cite{n100_2}, and the grid of explosion models explored do not extend to the fainter and faster evolving regime in which we find SN~2021rhu. More explosion models would be required to extend to these targets and we cannot, therefore, draw a strict conclusion on the suitability of SN~2021rhu for cosmology. We caution that its inclusion would also require the model light curve fitters \citep[e.g., SALT2;][]{SALT2guy}, to include transitional objects such as SN~2021rhu in their template libraries.

\section{Conclusion}
\label{conclusions}
SN~2021rhu is a transitional SN Ia bridging the gap between normal SNe Ia and those of the SN~1991bg-like class of underluminous thermounclear SNe.  Through abundance tomography we produced a custom abundance profile to describe the spectral evolution of SN~2021rhu using the N100 density profile. In order to reproduce the full velocity extent of the calcium features in the earliest spectra it was found to be necessary to extend this density profile up to 30000 \kms. We subsequently showed that this custom model - N100* - could be projected onto other extended density profiles to provide an equally valid description of the data. This shows that density profiles from different explosion mechanisms could potentially be used to produce the observed spectra, and as such this cannot be used alone to discriminate between explosion mechanisms with similar density profiles (e.g.~delayed-detonation Chandrasekhar mass and sub-Chandrasekhar double-detonation models). However, it was found that a density profile with densities lower than the N100, M0803, and M0905 profiles in the high-velocity regions is required to avoid unphysically high amounts of unburnt carbon. We also demonstrated that the W7 deflagration density profile was too low in the outer regions to produce the high velocity components of the observed calcium features. This also extends to the scaled W7 density profile which was chosen to be the preferred profile for SN~1986G in \cite{1986G_modelling}.

With the large scatter on the correlation between host extinction and diffuse interstellar band strength, the amount of extinction from the host galaxy in the case of SN~2021rhu is fairly unconstrained. We took the measurement of host extinction as an upper limit due to the fact that other similar transitional objects with this \TiII\ feature have all been subluminous, and as such we believe SN~2021rhu to sit below the Phillips relation. In order to compare the resulting models to those coming from the literature, it was required to look at the product of the abundance profile with the density profile being used. These species density profiles were in turn compared to literature models arising from the double-detonation mechanism and the delayed-detonation mechanism. This comparison demonstrated that while there are a couple of discrepancies between SN~2021rhu and the delayed-detonation models, namely the titanium bump, this transitional event is more consistent with a Chandrasekhar delayed-detonation than a sub-Chandrasekhar double-detonation. This suggests that SN~2021rhu may simply be a lower luminosity, and hence lower temperature, version of models that fit `normal' SNe Ia well and as such its use as H0 calibrator objects and in SN Ia samples for cosmological measurements is justified. Larger samples of this class of objects, with detailed abundance tomography, will allow for further investigation of the boundary between SNe Ia that sit on the Phillips relation and those that do not.

\section*{Acknowledgements}

The research conducted in this publication was funded by the Irish Research Council under grant number GOIPG/2020/1387. K.~M., M.~D., and G.~D. are funded by the EU H2020 ERC grant no. 758638. S.D. acknowledges support from the Marie Curie Individual Fellowship under grant ID 890695 and a junior research fellowship at Lucy Cavendish College. S.~S. acknowledges support from the G.R.E.A.T research environment, funded by {\em Vetenskapsr\aa det}, the Swedish Research Council, project number 2016-06012. S.R. acknowledges support by the Helmholtz Weizmann Research School on Multimessenger Astronomy, funded through the Initiative and Networking Fund of the Helmholtz Association, DESY, the Weizmann Institute, the Humboldt University of Berlin, and the University of Potsdam. M.S. and E.R. has received funding from the European Research Council (ERC) under the European Unions Horizon 2020 research and innovation program (grant agreement No.759194 - USNAC). M.~W.~C. acknowledges support from the National Science Foundation with grant numbers PHY-2010970 and OAC-2117997.

Based on observations obtained with the Samuel Oschin Telescope 48-inch and the 60-inch Telescope at the Palomar Observatory as part of the Zwicky Transient Facility project. Z.T.F. is supported by the National Science Foundation under grant No. AST-2034437 and a collaboration including Caltech, IPAC, the Weizmann Institute for Science, the Oskar Klein Center at Stockholm University, the University of Maryland, Deutsches Elektronen-Synchrotron and Humboldt University, the TANGO Consortium of Taiwan, the University of Wisconsin at Milwaukee, Trinity College Dublin, Lawrence Livermore National Laboratories, IN2P3, France, the University of Warwick, the University of Bochum, and Northwestern University. Operations are conducted by COO, IPAC,
and UW.

This research made use of \textsc{tardis}, a community-developed software
package for spectral synthesis in SNe
\citep{2014MNRAS.440..387K, kerzendorf_wolfgang_2020_3893940}.
The development of \textsc{tardis} received support from the
Google Summer of Code initiative
and from ESA's Summer of Code in Space program. \textsc{tardis} makes
extensive use of Astropy and PyNE.

SED Machine is based upon work supported by the National Science Foundation under Grant No. 1106171.

Based on observations made with the Nordic Optical Telescope, owned in collaboration by the University of Turku and Aarhus University, and operated jointly by Aarhus University, the University of Turku and the University of Oslo, representing Denmark, Finland and Norway, the University of Iceland and Stockholm University at the Observatorio del Roque de los Muchachos, La Palma, Spain, of the Instituto de Astrofisica de Canarias.

This work made use of the Heidelberg Supernova Model Archive (HESMA), \url{https://hesma.h-its.org}.

The image of the host galaxy NGC7814 along with SN~2021rhu was taken by the Chart32 team. Prompt 7 CTIO/UNC Chile, Chart32-Team.

\section*{Data Availability}
The observed spectra presented in this paper have been uploaded to WISeREP. The photometry can be found in the appendix below.


\bibliographystyle{mnras}
\bibliography{references} 

\appendix
\section{Data}

\begin{table*}
\caption{Photometric data for SN~2021rhu taken with the ZTF camera on the P48 telescope, as well as the \textit{w2}, \textit{m2}, \textit{w1}, \textit{u}, and \textit{v} bands taken with \textit{Swift}. These photometry measurements have not been corrected for extinction, and the ZTF measurements have been binned to a one day cadence. The \textit{Swift} measurments have been corrected for host contamination. The uncertainties are given in the brackets after the values as multiples of 0.01 mag.}
\parbox{.49\linewidth}{
\centering
\label{tab:photometry}
\begin{tabular}{cccc}
\hline
\textbf{MJD} & \textbf{ZTF~\textit{g}} & \textbf{ZTF~\textit{r}} & \textbf{ZTF~\textit{i}} \\ \hline
59396.4 & 15.73(2) & 15.61(2) &  \\
59399.4 &  &  & 15.24(8) \\
59400.4 & 14.39(2) & 13.8(2) &  \\
59402.4 & 13.38(2) & 13.21(2) & 13.42(2) \\
59403.4 & 13.14(2) & 12.9(2) & 13.1(2) \\
59404.4 & 12.93(2) &  &  \\
59405.4 &  & 12.59(2) & 12.82(2) \\
59406.4 &  & 12.59(2) &  \\
59407.4 &  & 12.31(2) & 12.71(2) \\
59408.4 &  & 12.4(2) & 12.72(2) \\
59409.4 & 12.42(2) & 12.33(2) &  \\
59410.4 & 12.38(2) & 12.32(2) &  \\
59411.4 & 12.39(2) & 12.43(2) & 12.78(2) \\
59412.4 & 12.43(2) & 12.32(2) &  \\
59414.4 & 12.49(2) & 12.37(3) &  \\
59415.4 &  & 12.4(2) &  \\
59417.4 & 12.65(2) & 12.56(2) & 13.22(2) \\
59418.4 & 12.76(2) & 12.49(2) &  \\
59422.4 & 13.13(2) & 12.95(2) & 13.39(2) \\
59423.4 & 13.24(2) & 12.98(2) & 13.37(2) \\
59424.4 & 13.5(2) & 13.01(2) & 13.29(2) \\
59425.4 & 13.63(2) & 13.09(2) &  \\
59426.4 & 13.66(2) & 13.04(2) &  \\
59427.4 & 13.78(2) & 13.06(2) & 13.18(2) \\
59428.4 & 13.86(2) & 13.06(2) &  \\
59429.4 & 14.15(2) & 13.17(2) & 13.15(2) \\
59430.4 &  & 13.22(2) & 13.21(2) \\
59434.4 &  & 13.55(2) & 13.43(2) \\
59435.4 & 14.71(2) & 13.73(2) & 13.52(2) \\
59436.4 & 14.74(2) & 13.74(2) & 13.59(2) \\
59438.4 & 14.86(2) & 13.91(2) &  \\
59439.4 & 14.98(2) & 13.99(2) & 13.82(2) \\
59440.4 & 15.0(2) &  &  \\
59442.4 & 15.06(2) &  & 14.03(2) \\
59443.4 & 15.06(2) & 14.27(2) &  \\
59444.4 & 15.13(2) &  &  \\
59446.4 & 15.2(2) & 14.43(2) & 14.31(2) \\
59447.4 &  &  & 14.34(2) \\
59448.4 & 15.27(2) &  &  \\
59449.4 & 15.22(2) &  & 14.39(2) \\
59450.4 & 15.29(2) & 14.52(2) & 14.55(2) \\
59451.4 &  & 14.59(2) &  \\
59452.4 & 15.32(2) & 14.68(2) &  \\
59453.4 & 15.39(2) & 14.66(2) & 14.63(2) \\
59454.4 & 15.48(2) & 14.67(2) & 14.62(2) \\
59455.4 &  &  & 14.72(2) \\
59456.4 & 15.45(2) &  &  \\
59458.4 &  & 14.85(2) & 14.85(2) \\
59459.4 & 15.53(2) & 14.91(2) & 14.9(2) \\
59460.4 & 15.56(2) & 14.96(2) & 15.04(2) \\

\hline
\end{tabular}

}
\hfill
\parbox{.49\linewidth}{
\centering
\begin{tabular}{cccc}
\hline
\textbf{MJD} & \textbf{ZTF~\textit{g}} & \textbf{ZTF~\textit{r}} & \textbf{ZTF~\textit{i}} \\ \hline
59461.4 & 15.58(2) & 14.96(2) & 15.03(2) \\
59462.4 & 15.64(2) &  & 15.09(2) \\
59463.4 & 15.61(2) & 15.04(2) &  \\
59464.4 & 15.61(2) & 15.08(2) & 15.16(2) \\
59465.4 & 15.67(2) & 15.11(2) & 15.16(2) \\
59467.4 & 15.67(2) & 15.17(2) &  \\
59468.4 &  &  & 15.34(2) \\
59469.4 &  & 15.25(2) &  \\
59471.4 &  & 15.35(2) & 15.5(2) \\
59474.4 & 15.8(2) & 15.43(2) & 15.63(2) \\
59476.4 & 15.84(2) & 15.49(2) &  \\
59477.4 &  &  & 15.71(2) \\
59478.4 & 15.88(2) & 15.59(2) &  \\
59480.4 & 15.9(2) & 15.63(2) & 15.87(2) \\
59484.4 &  &  & 15.91(2) \\
59485.4 & 16.0(2) & 15.83(2) &  \\
59487.4 & 16.04(2) & 15.91(2) & 16.13(2) \\
59489.4 & 16.06(2) & 15.97(2) &  \\
59491.4 &  &  & 16.22(2) \\
59494.4 & 16.17(2) & 16.18(2) & 16.38(2) \\
59497.4 &  &  & 16.52(2) \\
59498.4 & 16.22(2) &  &  \\
59502.4 & 16.36(2) & 16.46(2) & 16.64(2) \\
59504.4 & 16.32(2) & 16.39(2) &  \\
59505.4 &  &  & 16.73(2) \\
59509.4 &  &  & 16.76(2) \\
59512.4 & 16.46(2) & 16.67(2) & 16.9(2) \\
59517.4 &  &  & 17.02(2) \\
59518.4 & 16.62(2) &  &  \\
59520.4 &  &  & 17.14(3) \\
59521.4 & 16.67(2) & 16.96(2) &  \\
59523.4 & 16.77(2) &  & 17.22(3) \\
59525.4 & 16.82(2) & 17.02(2) &  \\
59526.4 &  &  & 17.25(3) \\
59527.4 & 16.78(2) & 17.1(2) &  \\
59529.4 &  &  & 17.43(4) \\
59530.4 &  & 17.19(2) &  \\
59532.4 & 16.86(2) &  & 17.41(3) \\
59537.4 & 16.95(2) & 17.42(3) &  \\
59540.4 & 17.02(2) & 17.57(3) &  \\
59542.4 &  &  & 17.72(5) \\
59550.4 &  &  & 17.72(5) \\
\hline
\end{tabular}

\parbox{\linewidth}{
\centering
\begin{tabular}{cccccc}
\hline
\textbf{MJD} & \textbf{\textit{w2}} & \textbf{\textit{m2}} & \textbf{\textit{w1}} & \textbf{\textit{u}} & \textbf{\textit{v}} \\ \hline
59410.8 & 17.22(4) & 17.77(5) & 15.85(3) & 13.95(3) & 12.46(3) \\
59413.5 & 17.41(3) & 17.90(4) & 16.04(3) & 14.15(3) & 12.44(3) \\
59417.9 & 17.86(4) & 18.34(5) & 16.49(4) & 14.63(6) &  \\
59420.1 & 18.20(8)& & & & \\
59426.1 & & & & & 13.15(7) \\ 
\hline
\end{tabular}
}

}
\end{table*}

\begin{figure*}
 \includegraphics[width = \linewidth]{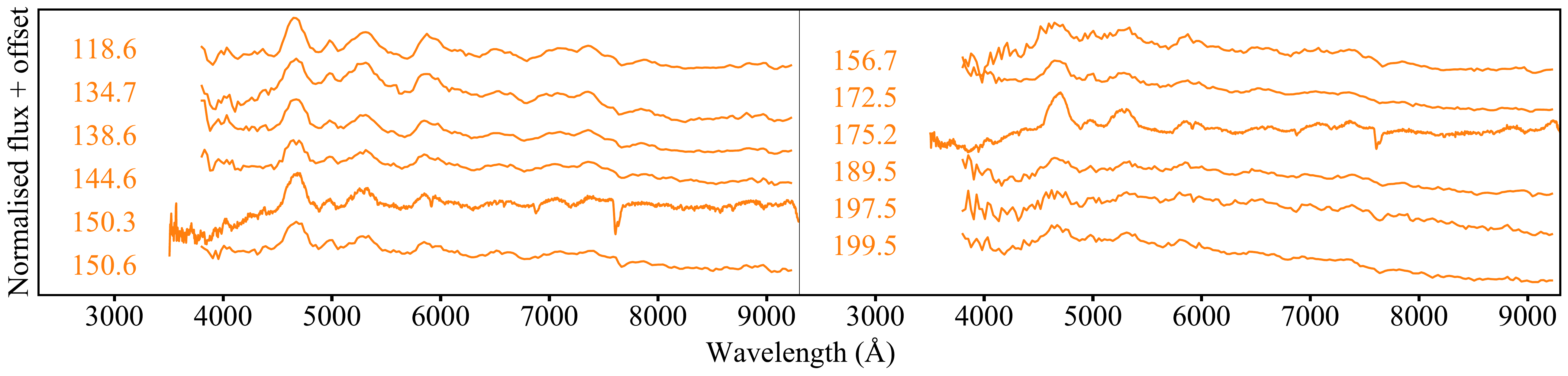}
 \caption{Remaining SN~2021rhu late time spectra with phases indicated alongside with respect to the date of peak brightness.}
 \label{fig:spectral_comp_appendix}
\end{figure*}

\begin{table*}
\caption{Observational details for the spectral data of SN~2021rhu. The wavelength ranges specified are given after correcting the spectra to the rest frame of the SN.}
\label{tab:spectral_info_2021rhu}
\begin{tabular}{ccccccc}
\hline
\textbf{MJD} & \textbf{Phase (d)} & \textbf{Telescope} & \textbf{Instrument} & \textbf{Wavelength range (Å)} \\ \hline
59398.37 & $-$12.2 & VLT & XShooter & 3089-24705 \\
59399.37 & $-$11.2 & VLT & XShooter & 3089-24705 \\
59400.37 & $-$10.2 & P60 & SEDM & 3789-9191 \\
59402.37 & $-$8.2 & VLT & XShooter & 3789-9191 \\
59403.17 & $-$7.4 & LT & SPRAT & 4015-7966 \\
59405.19 & $-$5.4 & LT & SPRAT & 4015-7966 \\
59409.15 & $-$1.4 & LT & SPRAT & 4015-7966 \\
59411.16 & 0.6 & LT & SPRAT & 4015-7966 \\
59416.18 & 5.6 & LT & SPRAT & 4015-7966 \\
59421.19 & 10.6 & LT & SPRAT & 4015-7966 \\
59423.09 & 12.5 & NOT & ALFOSC & 3492-9516 \\
59426.22 & 15.6 & LT & SPRAT & 4015-7966 \\
59434.10 & 23.5 & LT & SPRAT & 4015-7966 \\
59466.96 & 56.4 & LT & SPRAT & 4015-7966 \\
59467.21 & 56.6 & P60 & SEDM & 3789-9191 \\
59476.35 & 67.8 & P60 & SEDM & 3789-9191 \\
59487.41 & 76.8 & P60 & SEDM & 3789-9191 \\
59508.26 & 97.7 & P60 & SEDM & 3789-9191 \\
59529.22 & 118.6 & P60 & SEDM & 3789-9191 \\
59545.32 & 134.7 & P60 & SEDM & 3789-9191 \\
59549.20 & 138.6 & P60 & SEDM & 3789-9191 \\
59555.16 & 144.6 & P60 & SEDM & 3789-9191 \\
59560.89 & 150.3 & NOT & ALFOSC & 3492-9516 \\
59561.14 & 150.6 & P60 & SEDM & 3789-9191 \\
59567.24 & 156.7 & P60 & SEDM & 3789-9191 \\
59583.09 & 172.5 & P60 & SEDM & 3789-9191 \\
59585.82 & 175.2 & NOT & ALFOSC & 3492-9516 \\
59600.10 & 189.5 & P60 & SEDM & 3789-9191 \\
59608.11 & 197.5 & P60 & SEDM & 3789-9191 \\
59610.11 & 199.5 & P60 & SEDM & 3789-9191 \\\hline
\end{tabular}
\end{table*}

\begin{table*}
\caption{Correction values for the photometry and spectra of SN~2011fe, SN~2021rhu and SN~1986G, along with the Phillips relation parameter space measurements for those objects and SN~1991bg. It is noted that $R_v=3.1$ was used for the MW extinction correction and $R_v=2.57^{+0.23}_{-0.21}$ was used for the host correction of SN~1986G. \textit{a} : \protect\cite{IRSA_dustmap}, \textit{b} : \protect\cite{DIB_extinction}, \textit{c} : \protect\cite{1986G_distance_modulus}, \textit{d} : \protect\cite{Phillips99}, \textit{e} : \protect\cite{1986G_observations}, \textit{f} : \protect\cite{91bg_phillips}, \textit{g} : \protect\cite{2011fe_distance_modulus}, \textit{h} : \protect\cite{2011fe_photometry}, \textit{i} : \protect\cite{Suhail_H0}.}
\protect\label{tab:correction_values}
\begin{tabular}{ccccccc}
\hline
\textbf{Target} & \textbf{MW $A_\text{v}$} & \textbf{Host $A_\text{v}$} & \textbf{$\mu$} & $\Delta\textbf{m}_{\textbf{15}}$\textbf{(B)} & $\textbf{M}_{\textbf{B, peak}}$ \\ \hline
SN~1986G & $0.3054\pm0.0016$ $_\textit{a}$ & $2.03^{+0.09}_{-0.03}$ $_\textit{b}$ & $27.67 \pm0.12$ (Cepheid) $_\textit{c}$ & $1.73\pm0.07$ $_\textit{d}$ & $-18.24\pm0.13$ $_\textit{e}$\\ 
SN~1991bg & $0.0344\pm0.0016$ $_\textit{a}$ & 0 & - & $1.95\pm0.14$ $_\textit{f}$ & $-16.54\pm0.32$ $_\textit{f}$\\
SN~2011fe & $0.0239\pm0.0006$ $_\textit{a}$ & 0 & $29.04\pm0.05$ (Cepheid) $_\textit{g}$ & $1.21\pm0.03$ $_\textit{h}$ & $-19.21\pm0.15$ $_\textit{h}$\\
SN~2021rhu & $0.1187\pm0.0009$ $_\textit{a}$ & 0 / 0.42 &  $30.86\pm0.07$ (TRGB) $_\textit{i}$ & $1.52\pm0.2$ & $-18.37\pm0.2$ \\
\hline
\end{tabular}
\end{table*}

\begin{figure*}
 \includegraphics[width = \linewidth]{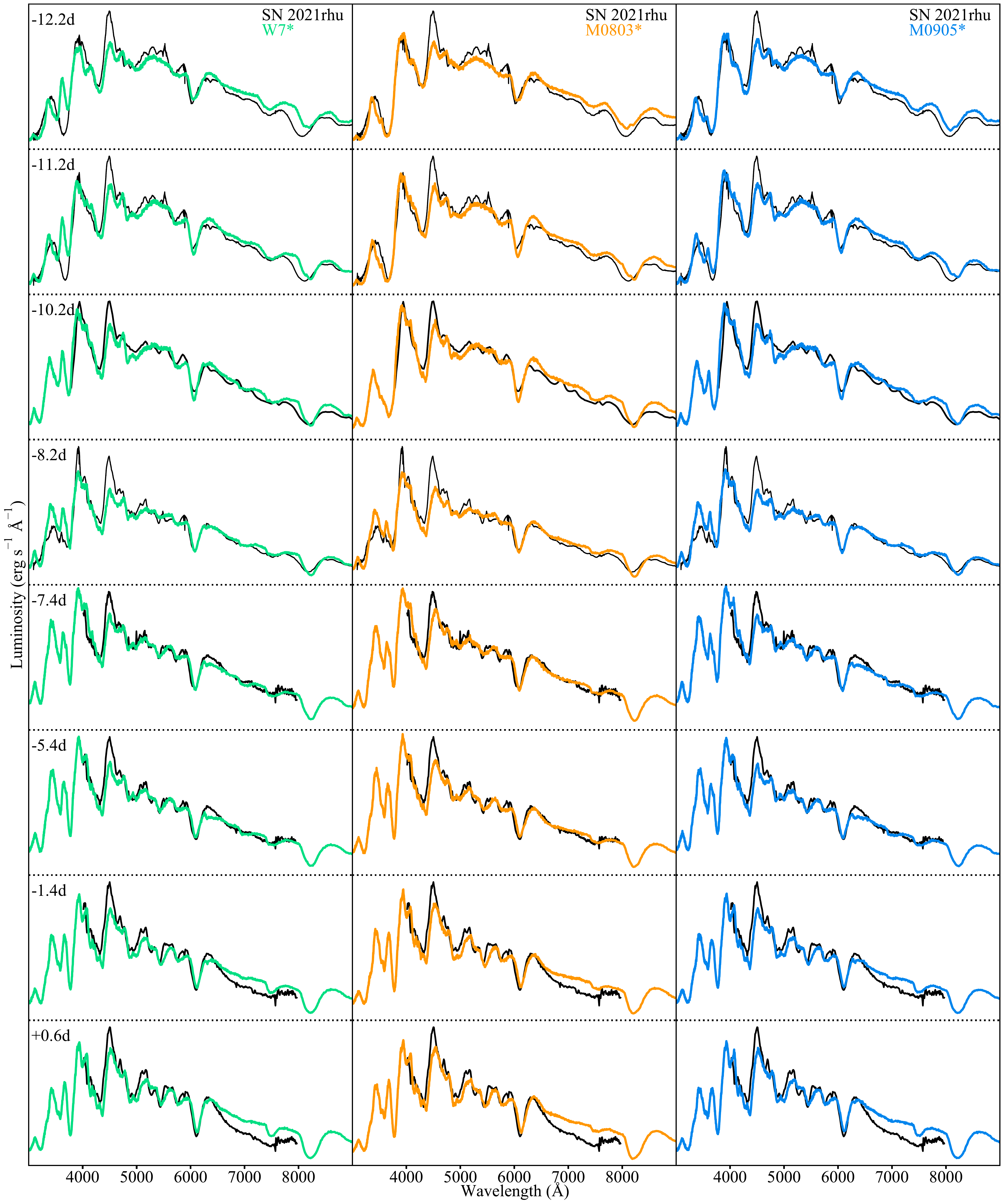}
 \caption{The synthesised spectral sequence produced from the models obtained through the projection of the N100* to the W7, M0803, and M0905 density profiles.}
 \label{fig:mangled_models}
\end{figure*}

\begin{figure*}
 \includegraphics[width = \linewidth]{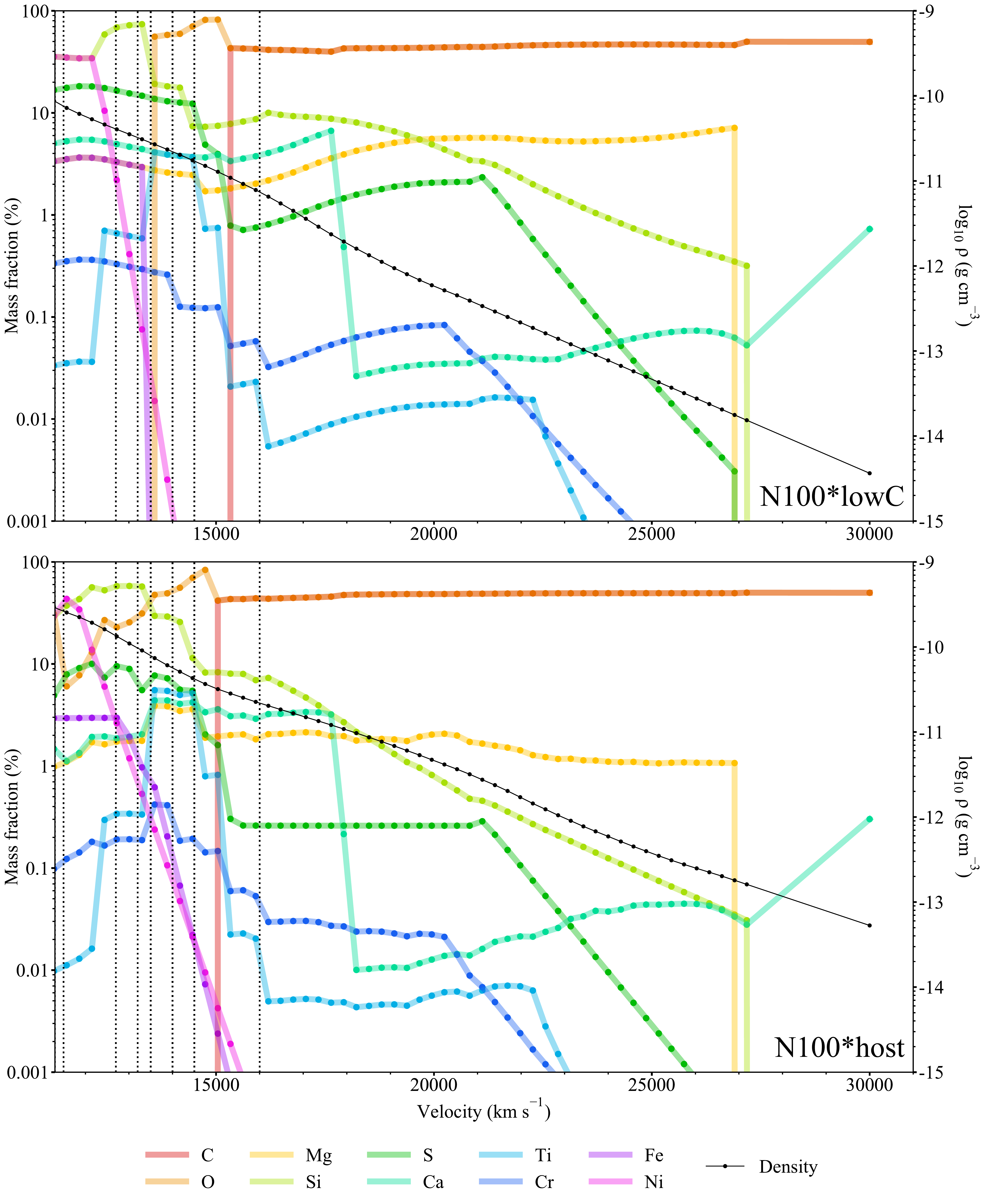}
 \caption{The preferred models for SN~2021rhu with (bottom) and without (top) host extinction corrections. The circles represent the shells and the vertical dotted lines indicate the photospheric velocities of the synthesised spectra. It is noted here that the photospheric velocities do not necessarily align exactly with the discrete shell velocities and as such the first shell in the model is taken as the first shell at this $v_{\text{ph}}$ or above. The density profiles overplotted in black are expanded to 1d post explosion. This is the N100 density profile in the bottom panel, and the 60 per cent kinetic energy N100 density profile in the top panel.}
 \label{fig:abundances_lowC_host}
\end{figure*}


\bsp	
\label{lastpage}
\end{document}